\documentclass[aps, prx, preprint, onecolumn, nofootinbib]{revtex4-2}
\usepackage{natbib}
\usepackage{import}
\usepackage{transparent}
\usepackage[colorlinks=true]{hyperref}
\usepackage{graphicx}
\usepackage{multirow}
\usepackage{verbatim, bm}
\usepackage{physics}
\usepackage{textcomp}
\usepackage{nicefrac}
\usepackage{mathrsfs}  
\usepackage[normalem]{ulem}
\usepackage{mathtools}
\usepackage{amsfonts}
\usepackage{amssymb}
\usepackage{color}
\usepackage{slashed}
\usepackage{amsmath}
\usepackage{wrapfig}
\usepackage{dsfont}
\usepackage{esint}
\hypersetup{
    unicode=false,          
    pdftoolbar=true,        
    pdfmenubar=true,        
    pdffitwindow=false,     
    pdfstartview={FitH},    
    pdfsubject={},          
    pdfcreator={},          
    pdfproducer={},         
    pdfkeywords={} {} {},   
    pdfnewwindow=true,      
    colorlinks=true,        
    linkcolor=magenta,      
    citecolor=blue,         
    filecolor=magenta,      
    urlcolor=blue           
} 
\allowdisplaybreaks

\renewcommand{\tilde}{\widetilde}

\renewcommand{\leq}{\leqslant}
\renewcommand{\geq}{\geqslant}
\renewcommand{\Re}{\operatorname{Re}}
\renewcommand{\Im}{\operatorname{Im}}

\begin{document}
\title{Conductance and thermopower fluctuations\texorpdfstring{\\}{} in interacting quantum dots}
\author{Henry Shackleton}

\author{Laurel E. Anderson}


\author{Philip Kim}


\author{Subir Sachdev}
\affiliation{Department of Physics, Harvard University, Cambridge MA 02138, USA}

\date{\today\vspace{0.4in}}

\begin{abstract}
    We model an interacting quantum dot of electrons by a Hamiltonian with random and all-to-all single particle hopping (of r.m.s. strength $t$) and two-particle interactions (of r.m.s. strength $J$). For $t \ll J$, such a model has a regime exhibiting the non-quasiparticle physics of the Sachdev-Ye-Kitaev model at temperatures $E_{\rm coh} \ll T \ll J$, and that of a renormalized Fermi liquid at $T \ll E_{\rm coh}$, where $E_{\rm coh} = t^2 / J$. Extending earlier work has computed the mean thermoelectric properties of such a dot weakly coupled to two external leads, we compute the sample-to-sample fluctuations in the conductance and thermopower of such a dot, and describe several distinct regimes. In all cases, the effect of the SYK interactions is to reduce the strength of the sample-to-sample fluctuations. We also find that in the regime where the mean transport co-efficients are determined only by the value of $J$ at leading order, the sample-to-sample fluctuations can be controlled by the influence of the smaller $t$.
\end{abstract}
\maketitle
\newpage 
\tableofcontents
\section{Introduction}
\label{sec:intro}

The Sachdev-Ye-Kitaev (SYK) model~\cite{sachdev1993,kitaev_talk} is a strongly interacting quantum many-body system without quasiparticle excitations, whose exact solvability in the large-$N$ limit - with $N$ the number of sites - has led to significant interest in it both as a toy model for non-Fermi liquid behavior as well as an analytically tractable example of holographic duality~\cite{chowdhury2022,kitaev2018}. 

In contrast to its analytic solvability, experimental realizations of the SYK model have proved to be challenging. The SYK model is defined microscopically as a system of fermions with random all-to-all quartic interactions, and is unstable at low temperatures to single-particle hopping. As such, any experimental proposal must generate strongly-disordered interactions with a high degree of connectivity, while simultaneously quenching any single-particle hopping terms. Several promising proposals have been made to this extent, involving Majorana zero modes~\cite{chew2017, pikulin2017}, quantum processors \cite{Garcia-Alvarez:2016wem,Babbush:2018mlj}, ultracold gases~\cite{danshita2017, Tigran21, Uhrich:2023ddx} and disordered graphene flakes~\cite{chen2018, brzezinska2022}. Simulations of the SYK model have been achieved on quantum processors~\cite{jafferis2022} and controllable nuclear-spin-chain simulators~\cite{luo2019}. Our study here was motivated by experiments on disordered graphene flakes \cite{LaurelThesis}, results of which will be reported in a separate paper \cite{Laurelpaper}.

Each experimental realization of the SYK model will have a different set of observables that it is best suited to study. Our focus will be on proposals for realizing the SYK model with complex fermions in a disordered graphene flake, for which the measurable quantities are thermoelectric transport observables, such as conductance and thermopower. Theoretical predictions for the average values of these quantities have been calculated~\cite{kruchkov2020} for realistic models that include both SYK terms and experimentally-relevant perturbations. The conclusion of this analysis is that thermoelectric quantities display a crossover from Fermi liquid-like behavior at temperatures below a coherence energy $E_{\text{coh}} = t^2/J$, where small single-particle hopping terms, with r.m.s. value $t$, produce coherent quasiparticle excitations, to SYK-like behavior at temperatures $E_{\rm coh} \ll T \ll J$, where $T$ is the temperature, and $J$ is the r.m.s. value of the SYK interactions.

In experimental realizations of these mesoscopic systems, transport quantities such as the conductance and thermopower will display sample-to-sample fluctuations, or alternatively fluctuations as a function of tuning external parameters such as chemical potential or magnetic field. For weakly-interacting disordered quantum dots at zero temperature coupled to broad multi-channel leads, this results in the well-studied phenomenon of \textit{universal conductance fluctuations} (UCF) at zero temperature, where the conductance displays $\order{1}$ fluctuations (in units of the conductance quanta, $e^2 / h$) whose magnitude is independent of the disorder strength~\cite{lee1985, altshuler1985, altshuler1986, lee1987}. An analogous treatment of disorder fluctuations in strongly-correlated quantum dots has not been explored previously. In this work, we analyze the fluctuations in transport properties in quantum dots with strong SYK interactions, and study the behavior of these fluctuations as their average values crossover from Fermi liquid-like for $T \ll E_\text{coh}$ to SYK-like for $T \gg E_\text{coh}$. We contrast analysis of these properties in the SYK regime, which involve statistical fluctuations of the \textit{single-particle} Green's function, with the large body of work analyzing statistical properties of the many-body spectrum~\cite{cotler2017,altland2018,saad2019,jia2020}.

Our analysis is able to recover UCF behavior for zero temperature, while the variance of the conductance in the Fermi liquid regime displays a $T^{-1}$ falloff at higher temperatures, consistent with prior studies of weakly-interacting disordered quantum dots~\cite{efetov1995}. However, we find a surprising feature of these fluctuations for temperatures larger than the coherence energy. In contrast to the mean values of transport quantities, whose behavior for $T \gg E_\text{coh}$ is well-described by a pure SYK model ($t=0$), the same is not true for the variance - at leading order in $N^{-1}$, the variance of the conductance for a pure SYK model is distinct from the variance in a model with SYK interactions and random hopping with r.m.s. value $t > \sqrt{TJ}/N$. The self-averaging properties of the pure SYK model are so strong that, to leading-order in $N^{-1}$, fluctuations of the physical transport properties remain driven by fluctuations of random hopping terms, even if their mean values are well-described by the pure SYK solution. Distinct predictions are still found for the two temperature regimes, arising from the different form of the average spectral function in the two limits, and we find a $T^{-2}$ suppression of the conductance variance in the SYK regime in contrast with the $T^{-1}$ Fermi liquid prediction.

These aspects of our results are illustrated by the following summary of our prediction for the mean $(\overline{\sigma})$ and variance of the electrical conductance $(\sigma)$:
\begin{align}
    \overline{\sigma}_{FF}  \propto \frac{\Gamma e^2}{\hbar} \frac{1}{t}\quad &, \quad \text{Var } \sigma_{FF} \propto \left(\frac{\Gamma e^2}{\hbar}\right)^2 \frac{1}{N t T} \label{res1} \\
    \overline{\sigma}_{SYK} \propto \frac{\Gamma e^2}{\hbar} \frac{1}{\sqrt{J T}} \quad &, \quad  \text{Var } \sigma_{SYK} \propto \left( \frac{\Gamma e^2}{\hbar} \right)^2 \frac{1}{N^3 JT} \label{res2} \\
    \overline{\sigma}_{tSYK}  \propto \frac{ \Gamma e^2}{\hbar} \frac{1}{t}\quad &, \quad \text{Var } \sigma_{tSYK} \propto   \left(\frac{\Gamma e^2}{\hbar}\right)^2 \frac{1}{N J T}, \quad T \ll E_{\rm coh} \label{res3} \\
    \overline{\sigma}_{tSYK} \propto \frac{\Gamma e^2}{\hbar} \frac{1}{\sqrt{J T}} \quad &, \quad \text{Var } \sigma_{tSYK} \propto  \left( \frac{\Gamma e^2}{\hbar} \right)^2 \frac{\mathcal{E}^2 t^2}{NJ^2 T^2} , \quad E_{\rm coh} \ll T \ll J \label{res4}
\end{align}
Here ({\it i\/}) $FF$ refers to the free-fermion results in 
Section~\ref{sec:conductance_statistics} with  $\Gamma$ a measure of the coupling to the leads, and the various of $\sigma_{FF}$ crosses over to the UCF value when $T< \Gamma^2/Nt$; ({\it ii\/}) the pure SYK results are in Section~\ref{sec:sykconductance}; ({\it iii\/}) $tSYK$ refers to the model with both hopping and interactions with $t \ll J$, $E_{\rm coh} = t^2/J$, the results for $T \ll E_{\rm coh}$ are in Section~\ref{sec:tSYKFL}, and the results for $ E_{\rm coh} \ll T \ll J$ are in Section~\ref{sec:tSYK} ($\mathcal{E}$ is a measure of the particle-hole asymmetry). All these results are obtained for the case where the coupling to the leads, $\Gamma$, is the smallest energy scale, and to leading order in a $1/N$ expansion.

Note that in all cases, the effect of the SYK interactions is to {\it reduce\/} the strength of the conductance fluctuations: \\
({\it i\/}) Eq.~\ref{res2} is suppressed by a factor of $1/N^3$ in contrast to $1/N$ in all other cases,\\ 
({\it ii\/}) Eq.~\ref{res3} is smaller than Eq.~\ref{res1} by a factor of $t/J$, and \\
({\it iii\/}) Eq.~\ref{res4} is smaller than Eq.~\ref{res3} by a factor of $E_{\rm coh}/T$.

The structure of this paper is as follows. In Section~\ref{sec:setup}, we make explicit the setup of our theoretical model as well as the assumptions used in calculating thermoelectric quantities. In Section~\ref{sec:freeFermion}, we calculate the fluctuations of transport quantities in the non-interacting limit, where properties are governed by single-particle random matrix theory (RMT). We emphasize that this approach is distinct from more standard approaches of modeling UCF phenomena using RMT~\cite{beenakker1997}, where calculations are done at zero temperature and involve the statistical treatment of transmission eigenvalues. Our treatment is primarily done at non-zero temperature and in the limit of weak environmental coupling, although we show that it is possible to extend our results down to zero temperature and recover $\order{1}$ universal fluctuations in an appropriate limit. In Section~\ref{sec:syk}, we study transport fluctuations in the SYK regime, presenting results both for pure SYK as well as more realistic models with random single-particle hopping. In Section~\ref{sec:syk24}, we study a model with both SYK and random hopping terms and demonstrate that the transport fluctuations for $T \gg E_{\text{coh}}$ are qualitatively different than that of a pure SYK model. 

In each of these sections, we discuss the fluctuations of the thermopower in addition to the conductance. The statistical properties of the thermopower require more care; in our formalism, we find that the thermopower is determined by a ratio of two Gaussian random variables, and hence the variance is formally not well-defined. An approximation to normality is still appropriate in certain parameter regimes for small fluctuations around the mean, and hence we can formally define a variance within this approximation. We state results given this assumption and find qualitatively similar behavior as the conductance variance, which is that the presence of strong SYK interactions serves to reduce the fluctuations around the mean value.
\section{Setup}
\label{sec:setup}
\subsection{Hamiltonian and transport coefficients}
Our goal is to characterize fluctuations in transport properties of disordered quantum dots with random all-to-all interactions. We model this quantum dot by the Hamiltonian
\begin{equation}
  \begin{aligned}
    H_{\text{dot}} &= \frac{1}{(2N)^{3/2}}\sum_{ij; kl} J_{ij ; kl} c_i^\dagger c_j^\dagger c_k c_l
  + \frac{1}{N^{1 / 2}}\sum_{ij} t_{ij} c_i^\dagger c_j - \mu \sum_i c_i^\dagger c_i
  \label{eq:sykHamiltonian}
  \end{aligned}
\end{equation}
where $J_{ij; kl}$ and $t_{ij}$ are complex random numbers with zero mean and variances $J^2$ and $t^2$, respectively. The complex SYK model is given by the first term, whereas the second term is a random single-particle hopping which leads to Fermi liquid behavior at low temperatures.

The quantum dot is coupled to two leads. Following the approach of~\cite{gnezdilov2018}, we model the leads by considering the Hamiltonian
\begin{equation}
  \begin{aligned}
    H = H_{\text{dot}} + \sum_{\vb{q}} \epsilon_{\vb{q}} a_{\vb{q}}^\dagger a_{\vb{q}} + \sum_{i, \vb{q}, \alpha} \left[\lambda_{i \alpha} c_i^\dagger a_{\vb{q} \alpha} + \lambda_{i\alpha}^* a_{\vb{q} \alpha}^\dagger c_i \right]\,.
    \label{eq:fullHamiltonian}
  \end{aligned}
\end{equation}
where $\alpha =  R\,, L$ labels the right and left leads. To parameterize the coupling to the leads, we define the matrices
\begin{equation}
  \begin{aligned}
  \Gamma_{ij}^\alpha = \pi \rho_{\text{lead}, \alpha}  \lambda_{i \alpha} \lambda_{j \alpha}^*\,,
  \end{aligned}
\end{equation}
where $\rho_{\text{lead}, \alpha}$ is the density of states in lead $\alpha$ near the Fermi level. We will assume $\rho_{\text{lead}, L} = \rho_{\text{lead}, R} \equiv \rho_{\text{lead}}$. 

We will find that the nature of the conductance fluctuations depends sensitively on how we model the coupling to the leads, $\lambda_{i\alpha}$. This is in contrast to the mean values of transport quantities, which is not as sensitive. We first make the assumption that the two couplings are proportional to each other, i.e. $\lambda_{iR} = \alpha \lambda_{iL}$ for some constant $\alpha$. With this constraint, it becomes possible to express transport properties solely in terms of the equilibrium Green's functions of the quantum dot. Using expressions derived in~\cite{costi2010}, we define
\begin{equation}
  \begin{aligned}
    \mathcal{L}_{ab} &= -\frac{i}{2 \pi \hbar} \int_{-\infty}^\infty \dd{\omega} \omega^{a + b - 2} f'(\omega) \Im \Tr \left[ \bm{\Gamma} \bm{G}^R\right] \,, \\
    \label{eq:transportFormula}
  \end{aligned}
\end{equation}
with $\bm{G}^{R, A}(\omega)$ the local retarded and advanced Green's function of $H$, both $N \times N$ matrices, $f(\omega)$ the Fermi function $f(\omega) = 1 / \left( e^{\omega / T} + 1 \right)$, and $\bm{\Gamma} \equiv \bm{\Gamma}^L \bm{\Gamma}^R / \left( \bm{\Gamma}^L + \bm{\Gamma}^R \right)$. For cases where the matrix $\bm{\Gamma}^L + \bm{\Gamma}^R$ is non-invertable, this equation is modified by omitting the null subspace of the matrix. The Green's functions must be solved for the full Hamiltonian, including the coupling to the leads; however, we will primarily be focused on the parameter regime where $\Gamma$ is the smallest energy scale and the Green's functions of the isolated system $H_{\text{dot}}$ are used.

The electric conductance $\sigma$, thermal conductance $\kappa$, and thermopower $\Theta$ are given by 
\begin{equation}
  \begin{aligned}
  \sigma = e^2 \mathcal{L}_{11}\,, \quad \kappa = \beta \left( \mathcal{L}_{22} - \frac{\mathcal{L}_{12}^2}{\mathcal{L}_{11}} \right)\,, \quad \Theta = \frac{\beta}{e} \frac{\mathcal{L}_{12}}{\mathcal{L}_{11}}\,. \\
  \label{eq:thermoelectricFormulas}
  \end{aligned}
\end{equation}
where $\beta = 1/T$.

Beyond this point, we must make further assumptions on the nature of the coupling to the leads. For notational simplicity, we will assume $\lambda_{iR} = \lambda_{iL} \equiv \lambda_i$ - generalization to the case where the magnitude of the couplings are asymmetric does not qualitatively affect our results.

\subsubsection{Single site lead coupling}
In this model, we take our two leads to be coupled to a single site, i.e. $\lambda_{i} \equiv \delta_{i 1} \lambda$. Defining $\Gamma \equiv \pi \rho_{\text{lead}} \abs{\lambda}^2$, we have
\begin{equation}
  \begin{aligned}
    \overline{\mathcal{L}_{ab}} &= \frac{2\Gamma}{ \pi \hbar} \int_{-\infty}^\infty \dd{\omega} \omega^{a + b - 2} f'(\omega) \overline{ \Im G^R_{11}(\omega) }
    \label{eq:transportFormula1}
  \end{aligned}
\end{equation}
Recall that the Green's functions are dependent on the random variables $J_{ij;kl
}, t_{ij}$. Averaging over disorder, we find that $\overline{\Im G^R_{11}(\omega)} = N^{-1} \sum_{ii} \overline{\Im G^R_{ii}(\omega)} \equiv \overline{\Im G^R(\omega)}$. Note that this relation relies on neglecting corrections to $\bm{G}^R$ arising from the couplings to the leads, as these corrections will be site-dependent.

Higher moments of these transport coefficients are given by
\begin{equation}
  \begin{aligned}
    \overline{\mathcal{L}_{ab} \mathcal{L}_{ab}} - \overline{\mathcal{L}_{ab}} \, \overline{\mathcal{L}}_{ab} &=  \left( \frac{2 \Gamma}{\pi \hbar} \right)^2 \int_{-\infty}^\infty \dd{\omega} \dd{\epsilon} \omega^{a+b-2} \epsilon^{a+b-2} f'(\omega) f'(\epsilon) \rho_d(\omega, \epsilon) \\
  \end{aligned}
\end{equation}
where we define
\begin{equation}
  \begin{aligned}
    \rho_d(\omega, \epsilon) &\equiv \frac{1}{N^2} \sum_{ij}\left[ \overline{\Im G^R_{ii}(\omega) \Im G^R_{jj}(\epsilon)} - \overline{\Im G^R_{ii}(\omega)} \, \overline{\Im G^R_{jj}(\epsilon)} \right]
  \end{aligned}
\end{equation}
The subscript $d$ indicates that this quantity describes the covariance of the \textit{diagonal} component of the Green's function, $G^R_{ii}$. 
\subsubsection{All-to-all couplings}
Here, we take the leads to be coupled to all sites with equal hopping, $\lambda_i \equiv \frac{\lambda}{\sqrt{N}}$. This model is also appropriate for hoppings that are equal in magnitude but with site-dependent phases, as the overall phase can be absorbed by a unitary transformation on the quantum dot operators. Defining $\Gamma \equiv \pi \rho_{\text{lead}} \abs{\lambda}^2$ as before, we have
\begin{equation}
  \begin{aligned}
    \overline{\mathcal{L}_{ab}} &= \frac{1}{N}\sum_{ij}\frac{2\Gamma}{ \pi \hbar} \int_{-\infty}^\infty \dd{\omega} \omega^{a + b - 2} f'(\omega)  \overline{\Im G^R_{ij}(\omega) } = \frac{2\Gamma}{ \pi \hbar} \int_{-\infty}^\infty \dd{\omega} \omega^{a + b - 2} f'(\omega)  \overline{\Im G^R(\omega)}\,.
  \end{aligned}
\end{equation}
where we utilize the fact that $\overline{G^R_{ij}(\omega)} = 0$ for $i \neq j$. The overall scaling of $N^{-\frac{1}{2}}$ in $\lambda_i$ was chosen such that the mean value of the conductance is consistent with the previous model. The second moment is given by
\begin{equation}
  \begin{aligned}
    \overline{\mathcal{L}_{ab} \mathcal{L}_{ab}} - \overline{\mathcal{L}_{ab}} \, \overline{\mathcal{L}_{ab}}&= \frac{1}{N^2}\sum_{ij, kl} \left( \frac{2\Gamma}{\pi \hbar} \right)^2 \int_{-\infty}^\infty \dd{\omega} \dd{\epsilon} \omega^{a+b-2} \epsilon^{a+b-2} f'(\omega) f'(\epsilon) \left[ \rho_d(\omega, \epsilon) + \rho_o(\omega, \epsilon) \right] \,. \\
  \end{aligned}
\end{equation}
where now we define the off-diagonal Green's function covariance,
\begin{equation}
  \begin{aligned}
    \rho_o(\omega, \epsilon) &\equiv \frac{1}{N^2} \sum_{ij}\left[ \overline{\Im G^R_{ij}(\omega) \Im G^R_{ji}(\epsilon)} - \overline{\Im G^R_{ij}(\omega)} \, \overline{\Im G^R_{ji}(\epsilon)} \right]\,.
    \label{eq:offDiagonalCovariance}
  \end{aligned}
\end{equation}
\subsubsection{Disordered all-to-all couplings}
If our sites physically correspond to spatially random modes, as is the case in graphene realizations of strongly interacting quantum dots in the zeroth Landau level, then it may be appropriate to model the coupling to the leads as additional random variables. To analyze this case, we treat $\lambda_i$ as Gaussian random variables:
\begin{equation}
  \begin{aligned}
    \overline{\lambda_{i}} &= 0\,, \\
    \overline{ \lambda_{i}^* \lambda_{j}} &=\delta_{ij} \frac{\lambda^2}{N}\,,
  \end{aligned}
\end{equation}
which in turn implies that $\overline{\Gamma_{ij}^\alpha} = \delta_{ij} \frac{\pi \rho_{\text{lead}}\lambda^2}{N} \equiv \delta_{ij} \frac{\Gamma}{N}$. Crucial to the calculation of fluctuations, we note the identity
\begin{equation}
  \begin{aligned}
    \overline{\Gamma_{ij}^\alpha \Gamma_{k l}^\beta}  = \left(\frac{\Gamma}{N} \right)^2 \left( \delta_{ij} \delta_{kl}  + \delta_{il} \delta_{j k}\right) \,.
  \end{aligned}
\end{equation}
The average values of the transport coefficients are the same as in the previous models.
Using the relation
\begin{equation}
  \begin{aligned}
    \overline{ (\Gamma_{ij}^R + \Gamma_{ij}^L)  (\Gamma_{k l}^R + \Gamma_{k l }^L)}  = \left( \frac{2 \Gamma}{N} \right)^2 \left( \delta_{ij} \delta_{kl} + \delta_{il}\delta_{j k} \right) \,,
  \end{aligned}
\end{equation}
we can obtain higher moments of the transport coefficients. This leads to a result for the variance almost identical to the uniform all-to-all couplings in the previous section. The crucial difference is that in this case, the disconnected component of $\rho_o(\omega, \epsilon)$, defined in Eq.~\ref{eq:offDiagonalCovariance}, is not subtracted off in the expression for the variance of $\mathcal{L}_{ab}$. The consequence of this is a trivial contribution to the variance of $\mathcal{L}_{ab}$, which is given by $N^{-1} \overline{\mathcal{L}_{ab}}^2$ and can be thought of as being driven by the disorder in the leads in contrast to the intrinsic disorder in the quantum dot. While this is suppressed by a factor of $N^{-1}$, we will find that fluctuations generically only appear at the order or higher, so this contribution cannot be disregarded on these grounds.

We have shown that the variance of transport quantities, such as the conductance, are determined by the single-particle Green's function covariances $\rho_d$, $\rho_o$. The primary focus of our paper will be an analysis of these functions, and their implications for conductance fluctuations. For concreteness, we will give our predictions for conductance fluctuations in a model with uniform all-to-all couplings, such that both $\rho_o$ and $\rho_d$ contribute, and so that there is no trivial contribution to the variance arising from disordered leads. We summarize results for single-mode couplings in Appendix~\ref{app:singleLead}.
\subsection{Comparison to other analyses}
Due to the extensive literature on conductance fluctuations in mesoscopic systems, we make precise here the connection between our setup and prior work. 

The most well-established results for conductance fluctuations pertain to the $T=0$ conductance of a sufficiently weakly-interacting quantum dot such that a single-particle picture is appropriate. In this limit, conductance fluctuations can be understood most directly via a random matrix analysis of the scattering matrices, which take values in the circular ensemble~\cite{baranger1994, jalabert1994}. An alternative approach, suitable for studying the effects of non-zero temperature and weak magnetic fields, is to start with a microscopic single-particle Hamiltonian modeled as a random matrix, much like our Hamiltonian in the limit $J = 0$. In the non-interacting limit, the conductance for a generic set of lead couplings $\lambda_{jL}$, $\lambda_{jR}$ is given by the Landauer formula for a single channel,
\begin{equation}
  \begin{aligned}
    \sigma &= \frac{e^2}{h} \int \dd{\omega} f'(\omega) t(\omega) t^*(\omega) \,,
    \\
    t(\omega) &\equiv 2\pi \rho_{\text{lead}}\sum_{i j} \lambda_{iL}^* G^R_{ij}(\omega) \lambda_{jR}\,.
  \end{aligned}
\end{equation}
The conductance variance is thus related to the disorder average of four copies of $G^R$, solved in the presence of the leads. This becomes a tractable problem in the limit where the number of channels in the leads is large, and can be dealt with rigorously using supersymmetry techniques~\cite{verbaarschot1985a, iida1990, iida1990a, efetov1995, frahm1995} to give results consistent with random matrix predictions. This formulation can also be generalized to non-zero temperature~\cite{efetov1995}, where a $T^{-1}$ falloff of the conductance variance is observed. However, these supersymmetry techniques cannot be generalized to accommodate strong interactions.
Our results are most appropriately compared to the prior results on closed quantum dots, where the number of channels is small and are weakly coupled to the dot. These works have primarily focused on the $T=0$ conductance in either the weakly-interacting limit~\cite{mccann1996, prigodin1993}, where non-Gaussian behavior of the conductance was found, or in the Coulomb blockade-dominated limit~\cite{jalabert1994} which found a non-Gaussian distribution of the conductance peaks. To our knowledge, conductance fluctuations in the parameter regime of closed quantum dots with $T\gg \Gamma$ and for negligible Coulomb blockade effects has not been studied previously. This is the regime where we will conduct our analysis, as it is in this limit that the effects of strong SYK interactions becomes analytically tractable.

\section{Free fermion analysis}
\label{sec:freeFermion}
\subsection{Conductance statistics}
\label{sec:conductance_statistics}
We begin with an analysis of conductance fluctuations in the non-interacting limit ($J=0$). In this limit, the conductance is independent of temperature~\cite{kruchkov2020},
\begin{equation}
  \begin{aligned}
  \overline{\sigma} = \frac{e^2}{\hbar} \frac{\Gamma \sqrt{4t^2 - \mu^2}}{\pi t^2}\,.
  \label{eq:freeFermionConductanceMean}
  \end{aligned}
\end{equation}
In order to understand the behavior of conductance flucutations, we must calculate the single-particle covariances $\rho_{d, o}(\omega, \epsilon)$. This may be done diagrammatically, only keeping diagrams to leading order in $N^{-1}$. We do this by calculating the covariance of the Green's function in imaginary time and analytically continuing to the real axis. The calculation of $\rho_d$ involves analytic continuation of the quantity $\sum_{ij} G_{ii}(i \omega) G_{jj}(i \epsilon)$, and for $\rho_o$, $\sum_{ij} G_{ij}(i \omega) G_{ji}(i\epsilon)$. 

\begin{figure}[htpb]
  \centering
  \includegraphics[width=0.6\textwidth]{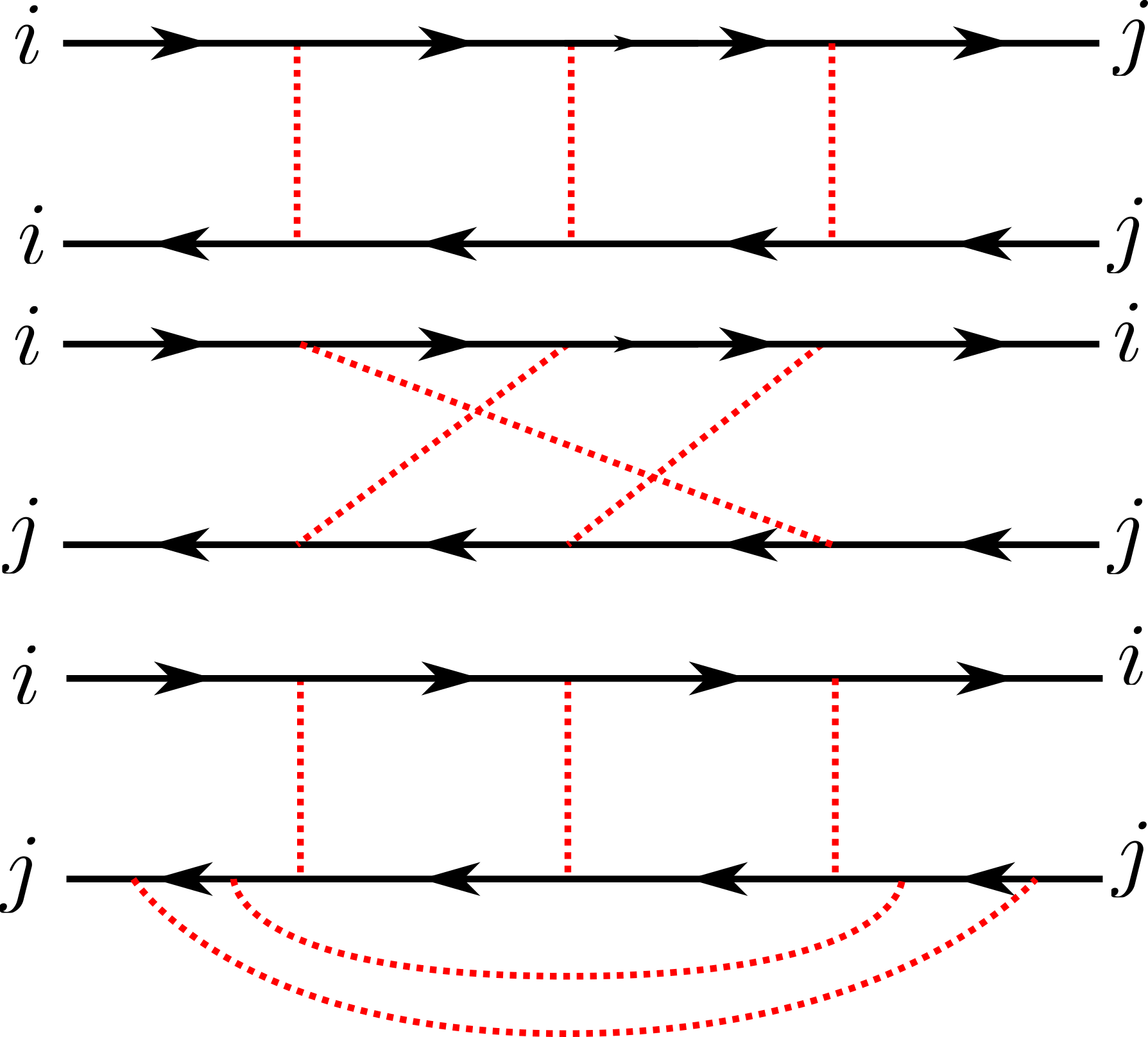}
  \caption{Ladder diagrams that contribute to the fluctuations of the single-particle spectral function. The first class of diagrams contributes to both the covariances $\rho_d$ and $\rho_o$, with the contribution to $\rho_d$ coming from the $i=j$ case. The last two classes only contribute to $\rho_d$. Disorder-averaging of the single-particle hopping (SYK interactions) is represented in red (blue). }
  \label{fig:syk2ladders}
\end{figure}

Diagrams that contribute to the covariance of the Green's function consists of diagrams of pairs of Green's functions that are only connected along disorder lines. The structure of these diagrams is shown in Fig.~\ref{fig:syk2ladders}. The diagrammatic structure of both the $\rho_d$ and $\rho_o$ fluctuations are similar - both involve an infinite summation over a set of ladder diagrams, given in the first figure in Fig.~\ref{fig:syk2ladders}. The leading order contributions to $\rho_o$ are just given by this set of diagrams. For $\rho_d$, two additional classes of diagrams must be considered and are shown in Fig.~\ref{fig:syk2ladders}. The first class yields an $n$-fold degeneracy of ladders with $n$ rungs, and the second class gives additional disorder averaging on either side of the ladder rungs. 

Putting all this together, we obtain the final form for the Green's function covariances,
\begin{equation}
  \begin{aligned}
g_d(i \omega, i \epsilon) &\equiv  \frac{1}{N^2} \sum_{ij}\left(  \overline{G_{ii}(i \omega) G_{jj}(i \epsilon)} - \overline{G_{ii}(i \omega)} \times \overline{ G_{jj}(i \epsilon)} \right) 
\\
&= \frac{1}{N^2} \frac{t^2 G(i\omega)^2 G(i\epsilon)^2}{\left[ 1 - t^2 G(i \omega) G(i\epsilon) \right]^2 } \frac{1}{1 - t^2 G(i\omega)^2}\frac{1}{1 - t^2 G(i\epsilon)^2}
  \\
  g_o(i\omega, i\epsilon) &\equiv \frac{1}{N^2} \sum_{ij} \left(\overline{G_{ij}(i \omega) G_{ji}(i \epsilon)} - \overline{G_{ij}(i \omega)} \times \overline{ G_{ji}(i \epsilon)} \right)  = \frac{1}{N} \frac{t^2 G(i\omega)^2 G(i\epsilon)^2}{ 1 - t^2 G(i \omega) G(i\epsilon) } 
  \label{eq:syk2ladders}
  \end{aligned}
\end{equation}
where in the RHS, we use the average Green's function
\begin{equation}
  \begin{aligned}
    G_0(i \omega) = \frac{i \omega + \mu}{2t^2} - i \frac{\text{sgn}(\omega)}{2t^2}\sqrt{4t^2 + (\omega - i \mu)^2}\,.
  \end{aligned}
\end{equation}
To obtain expressions for $\rho_{o, d}$, we analytically continue these to the real axis,
\begin{equation}
  \begin{aligned}
    \rho_\alpha = -\frac{1}{4} \left[ g_\alpha(\omega^+, \epsilon^+) + g_\alpha(\omega^-, \epsilon^-) - g_\alpha(\omega^-, \epsilon^+) - g_\alpha(\omega^+, \epsilon^-)   \right]
  \end{aligned}
\end{equation}
where $\omega^\pm \equiv \omega \pm i \eta$, $\eta \rightarrow 0$. The expression for $\rho_d$ has been derived before using a similar diagrammatic approach~\cite{brezin1994}, although we are not aware of an analogous calculation for $\rho_o$. 

From this analysis, we see that fluctuations arising from $\rho_o$ are enhanced relative to the $\rho_d$ fluctuations by a factor of $N$, and hence will be the main focus of our analysis. However, we will show that a more careful analysis of $\rho_d$ will be necessary to recover UCF behavior at zero temperature.

Due to the form of the average Green's function, we find a singular behavior for the Green's function covariances in Eq.~\ref{eq:syk2ladders} for $\abs{\omega - \epsilon} \rightarrow 0$, as
\begin{equation}
  \begin{aligned}
    1 - t^2 G(\omega^+) G(\epsilon^-) &= \frac{1}{t} \left(- \eta  +  \frac{i}{2}(\omega - \epsilon)\right) + \order{ (\omega/t)^2, (\epsilon / t)^2}\,,
    \\
    \rho_d(\omega, \epsilon) &= -\frac{1}{8 N^2} \Re \left[ \frac{1}{ (i(\omega - \epsilon)/2 - \eta )^2}  \right]
    \\
    \rho_o(\omega, \epsilon) &= -\frac{1}{2Nt} \Re \left[ \frac{1}{i (\omega - \epsilon) / 2 - \eta }   \right]  \\
    \label{eq:divergences}
  \end{aligned}
\end{equation}
The above divergence holds for arbitrary chemical potential $\mu$. We see that the $(\omega - \epsilon)^{-2}$ divergence in $\rho_d(\omega, \epsilon)$ is \textit{independent} of the energy scale $t$. The correlation function $\rho_d$ determines fluctuations of the single-particle energy levels - for the non-interacting system, the distribution of single-particle energy levels is determined by the Gaussian Unitary Ensemble (GUE) in which fluctuations are known to take this universal form~\cite{dyson1963, dyson1972, brezin1994}.

For $T\neq 0$, this divergence may be regulated by carefully taking the $\eta \rightarrow 0$ limit in the analytic continuation to the real axis. We state the calculation in a general form, for use later. For real-valued functions $A(\omega)$, $B(\omega)$, and $\rho(\omega-\epsilon) = \rho(\epsilon-\omega)$, we have the identity
\begin{equation}
\int \dd{\omega} \dd{\epsilon} A(\omega) B(\epsilon) \rho(\omega - \epsilon) = \sqrt{2\pi}  \int  \dd{k}\Re\left[\tilde{A}(k) \tilde{B}^*(k)\right] \tilde{\rho}(k)\,.
\label{eq:linearStatisticFormula}
  \end{equation}
where we define the Fourier transform $\tilde{A}(k) \equiv \frac{1}{\sqrt{2\pi}}\int e^{- i k x} A(x) \dd{x}$.

The Fourier transform of the Green's function covariances are:
\begin{equation}
    \begin{aligned}
        \tilde{\rho}_d(k) &= \frac{1}{8N^2} \sqrt{\frac{\pi}{2}} \abs{k}\,,
        \\
        \tilde{\rho}_o(k) &= \frac{\sqrt{2\pi}}{4 N t}\,.
    \end{aligned}
\end{equation} 
This analysis for $\rho_d$ recovers the well-known Dyson-Mehta formula for the variances of linear statistics in RMT~\cite{dyson1972, dyson1963}, and the more general covariance formula for linear statistics~\cite{cunden2014}; however, these fluctuations are a factor of $N^{-1}$ smaller than the contributions from $\rho_o$ fluctuations.

\begin{figure}[htpb]
  \centering
  \includegraphics[width=0.9\textwidth]{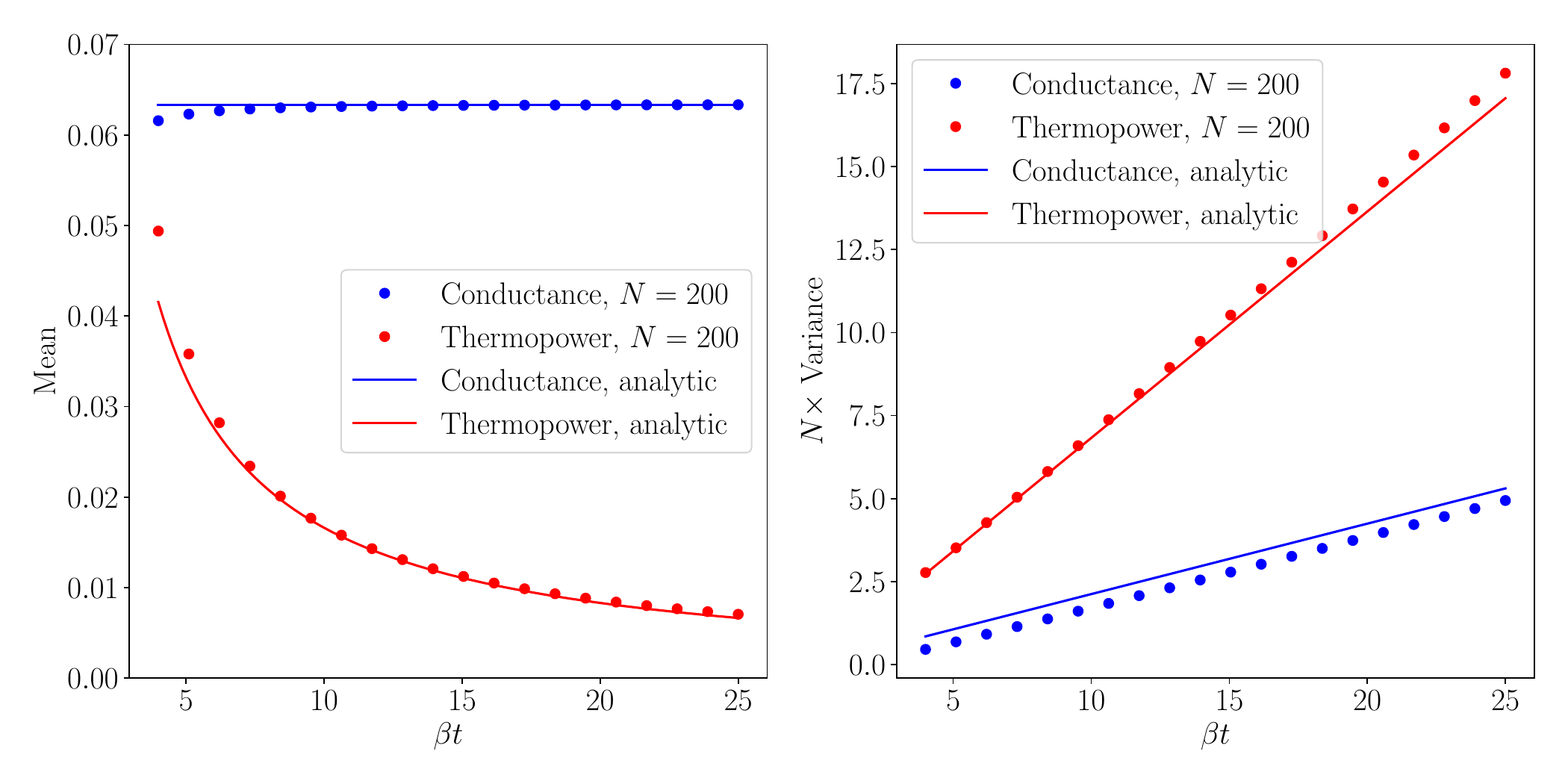}
  \caption{We plot the mean and variance of both the conductance and thermopower, calculated in the non-interacting ($J=0$) limit of our disordered quantum dot and using Eq.~\ref{eq:transportFormula} averaged over $100000$ realizations of the hoppings $t_{ij}$. We set the chemical potential $\mu = 0.33$. In this calculation, the Green's function of the quantum dot is solved independent of the leads. We set the strength of the leads coupling $\Gamma = 0.1$ in order for the mean thermopower and conductance to have comparable magnitudes, although we emphasize that this value only appears as an overall coefficient in the conductance. These numerical results are compared with the analytic predictions given in Eq.~\ref{eq:syk2conductanceVariance} and Eq.~\ref{eq:syk2ThermopowerVariance}, which show good agreement.}
  \label{fig:RMTPred}
\end{figure}

This formula yields the result for the conductance variance,
\begin{equation}
  \begin{aligned}
    \text{Var } \sigma =\left(\frac{\Gamma e^2}{\hbar}\right)^2 \frac{2}{3 \pi t T N}\,,
    \label{eq:syk2conductanceVariance}
  \end{aligned}
\end{equation}
which agrees well with a numerical simulation, shown in Fig.~\ref{fig:RMTPred}. This expression is valid for $\Gamma^2 \ll NTt$, as suggested by the $T \rightarrow 0$ divergence. In order to obtain results for $T=0$, a more careful treatment of the coupling to the leads is required.

To obtain a $T\rightarrow 0 $ result, we must include the self-energy arising from the coupling to the leads. The form of this correction is dependent on the manner in which we choose the coupling. For uniform all-to-all couplings, we have
\begin{equation}
  \begin{aligned}
    \Sigma_{ij}(i \omega) = 2 \int \dd{\epsilon} \frac{\rho(\epsilon) \abs{\lambda}^2}{i \omega - \epsilon} \approx \frac{2 i\Gamma}{N} \text{sgn}(\omega)
  \end{aligned}
\end{equation}
where $\rho$ is the density of states in the leads, which we approximate by its value at the Fermi level. To leading order in $\Gamma$,
\begin{equation}
  \begin{aligned}
    \overline{G_{ij}(i \omega)} = \delta_{ij} G_0(i\omega) + \frac{ 2 i\Gamma G_0(i\omega)^2}{N} \text{sgn}(\omega)
  \end{aligned}
\end{equation}
Using this result for $G$ in Eq.~\ref{eq:syk2ladders}, we can directly evaluate the $T=0$ conductance fluctuations
\begin{equation}
  \begin{aligned}
    \Gamma^2 \rho_d(0, 0) &= -\frac{1}{32} + \order{\Gamma / t}\,,
    \\
    \Gamma^2 \rho_o(0, 0) &= \order{\Gamma / t} \,,
    \\
    \text{Var } \sigma(T=0) &= \left(\frac{e^2}{h}\right)^2 \left( \frac{1}{8} + \order{\Gamma / t} \right) \,.
  \end{aligned}
\end{equation}
Note that in this case, $\rho_d$ and $\rho_o$ contribute at the same order; the $T=0$ divergences are regulated by an $\order{N^{-1}}$ self-energy, and $\rho_d$ is more singular at $T=0$. We stress that this result is not rigorous - as evident from the above results, this manner of including the corrections from the leads is not done consistently in an $N^{-1}$ expansion. A proper extrapolation down to $T=0$ necessitates, for example, the use of supersymmetric techniques~\cite{iida1990}.
\subsection{Thermopower statistics}
\label{sec:thermopower}
Although less well-studied than conductance fluctuations, thermopower fluctuations have been studied analytically for single-mode contacts at $T=0$~\cite{vanlangen1998} and for broad contacts~\cite{esposito1987}. Experimental measurements~\cite{gallagher1990, godijn1999} have found good agreement with these predictions. Our analysis will fall in a distinct parameter regime to these results, where we consider a quantum dot weakly coupled to its environment, at temperatures much larger than the coupling strength.

In the free fermion limit, the mean thermopower vanishes linearly with temperature~\cite{kruchkov2020}
\begin{equation}
  \begin{aligned}
    \overline{\Theta} = \frac{\pi^2 T}{3e} \frac{\mu}{4t^2 - \mu^2}\,.
  \end{aligned}
  \label{Thetalinear}
\end{equation}
The linear temperature dependence is a consequence of the linear temperature dependence of the entropy, and hence is generic for systems with quasiparticle excitations.

In our framework, the statistical properties of the thermopower is determined by the \textit{ratio} of two random variables, $\mathcal{L}_{12}$ and $\mathcal{L}_{11}$. As higher order moments are suppressed by additional factors of $N^{-1}$, our transport coefficients are Gaussian to leading order in $N^{-1}$. The thermopower distribution is then determined by the ratio of two Gaussian statistics, which in general is non-Gaussian. Nevertheless, an approximation to Gaussian is appropriate~\cite{hinkley1969} for capturing small fluctuations around the mean value, so long as the width of the Gaussian distribution is small relative to the mean. We provide more details on this approximation in Appendix~\ref{sec:ratioApp}. A similar approach was used to characterize fluctuations of the Fano factor in weakly-interacting quantum dots~\cite{cunden2015}. Such an approximation requires knowledge of the covariance of the two quantities $\mathcal{L}_{12}$ and $\mathcal{L}_{11}$. This is given by
\begin{equation}
  \begin{aligned}
    \text{Cov}(\mathcal{L}_{11}, \mathcal{L}_{12}) &= - \left( \frac{\Gamma}{ \pi \hbar} \right)^2 \int \dd{\omega} \dd{\epsilon} \omega f'(\omega) f'(\epsilon)  \left[ \rho_d(\omega, \epsilon) + \rho_o(\omega, \epsilon) \right]
  \end{aligned}
\end{equation}
which vanishes by an application of Eq.~\ref{eq:linearStatisticFormula}.
Therefore to leading order in $N^{-1}$, the random variables $\mathcal{L}_{11}$ and $\mathcal{L}_{12}$ are both uncorrelated and have a bivariate normal distribution, so we treat them as independent. With this assumption, the typical fluctuations of $\Theta$ around its mean value $\frac{\beta}{e} \frac{\overline{\mathcal{L}_{12}}}{\overline{\mathcal{L}_{11}}}$ are Gaussian with variance
\begin{equation}
  \begin{aligned}
    \frac{\text{Var } \Theta}{\overline{\Theta}^2} = \frac{\text{Var } \mathcal{L}_{11}}{\overline{\mathcal{L}_{11}}^2} + \frac{\text{Var } \mathcal{L}_{12}}{\overline{\mathcal{L}_{12}}^2}\,,
    \label{eq:ratioStatistic}
  \end{aligned}
\end{equation}
with
\begin{equation}
  \begin{aligned}
    \text{Var } \mathcal{L}_{11} &= \left( \frac{\Gamma}{\hbar} \right)^2  \frac{2 }{3 \pi N  T t} \quad \overline{\mathcal{L}}_{11} = \frac{\Gamma}{\hbar}\frac{\sqrt{4 t^2 - \mu^2}}{2 \pi t^2}   \\
    \text{Var } \mathcal{L}_{12} &= \left( \frac{\Gamma}{\hbar} \right)^2 \frac{(\pi^2 - 6) T}{9  \pi N t}   \quad \overline{\mathcal{L}_{12}} = -\frac{\Gamma}{\hbar} \frac{\pi \mu T^2}{6  t^2 \sqrt{4 t^2 - \mu^2}}\,. \\
    \label{eq:syk2ThermopowerVariance}
  \end{aligned}
\end{equation}
This analytic prediction agrees well with the numerically calculated variance, shown in Fig.~\ref{fig:RMTPred}. Similar to the conductance variance, the thermopower variance scales as $T^{-1}$ at low temperature, although the fact that the mean value scales linearly with temperature means that, in contrast to the conductance, the variance normalized by the mean squared diverges as $T^{-3}$.

\section{Pure SYK analysis}
\label{sec:syk}
\subsection{Conductance statistics}
\label{sec:sykconductance}
We now move to an analysis of conductance fluctuations for a pure SYK model ($t = 0$), where the average value takes the form at half-filling
\begin{equation}
  \begin{aligned}
      \overline{\sigma} &= \frac{e^2}{\hbar} \frac{0.72 \Gamma}{\sqrt{J T}}
      \label{eq:sykConductanceMean}
  \end{aligned}
\end{equation}
(the exact value of the prefactor is $2 \sqrt{2} \pi^{-1/4} \Gamma(3/4) \Gamma(1/4)\approx 0.72$).
Deviations away from half-filling only constitute a change in the numerical coefficient. For full generality, we present results for an $\text{SYK}_q$ model with $q$-fermion interactions - the case $q=4$ is the one of experimental relevance. The diagrammatic prescription for calculating the Green's function covariances $\rho_d$, $\rho_o$ remain the same, and we consider pairs of Green's functions that are only connected via disordered lines. The $N$ scaling of disorder-connected diagrams has been considered in SYK-like models previously~\cite{kitaev2018, gu2020, wang2019, shi2023}, although an explicit evaluation of such diagrams has only been carried out for the off-diagonal covariance, $\rho_o$, in the Majorana SYK model~\cite{wang2019}.

\begin{figure}[htpb]
  \centering
  \includegraphics[width=0.6\textwidth]{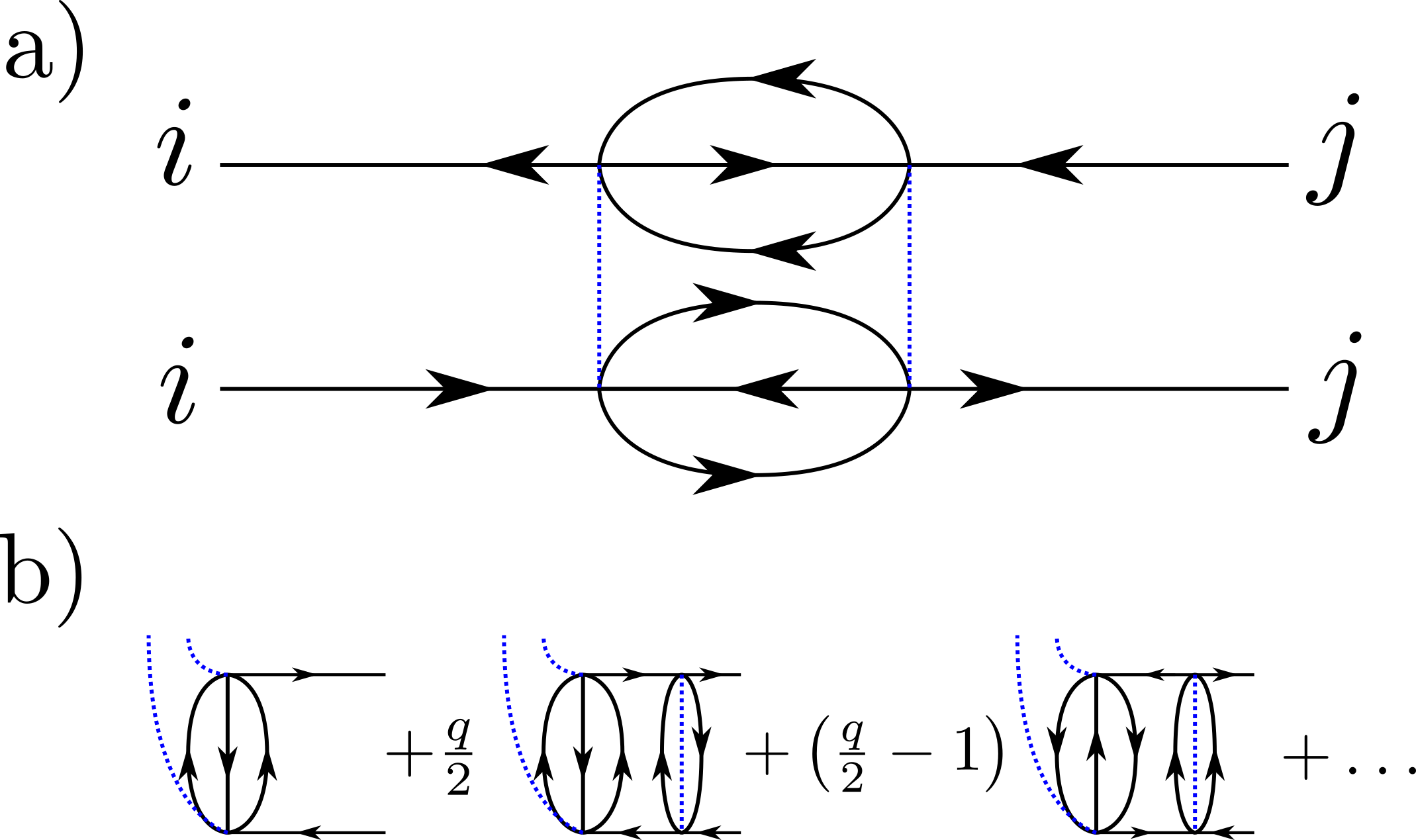}
  \caption{a) The leading-order diagram for a pure SYK model that contributes to the Green's function covariance. This contributes to $\rho_o$ with a factor of $N^{1-q}$, and the specialized $i=j$ case contributes to $\rho_d$ with a factor of $N^{-q}$. b) For the diagonal covariance $\rho_d$, the diagram in a) is the first in an infinite series of diagrams, generated from the first by attaching ladder rungs to either the top or bottom diagram in the manner shown here. We have deformed the diagram from a) in order to more clearly illustrate the structure of the ladder rungs.}
  \label{fig:complexSYKConnected}
\end{figure}

The simplest leading-order diagram, which contributes to both $\rho_d$ and $\rho_o$, is shown on top in Fig.~\ref{fig:complexSYKConnected}. This contributes to $\rho_o$ with coefficient $N^{1-q}$ and $\rho_d$ with coefficient $N^{-q}$. The different coefficients arise because the $\rho_d$ contribution appears with a factor of $\delta_{ij}$. We find that $\rho_d$ contains additional ``ladder'' diagrams as shown in the bottom of Fig.~\ref{fig:complexSYKConnected} that also contribute at $\order{N^{-q}}$.

Both these covariances can be evaluated analytically in the conformal limit, when $\beta J \gg 1$. To see this, we examine the first diagram, in the top of Fig.~\ref{fig:complexSYKConnected}. In the conformal limit, the Green's functions take the following form:
\begin{equation}
    g^R(\overline{\omega}, \overline{T}) = - i e^{-i \theta} \left( \frac{\pi}{\cos 2 \theta} \right)^{1/4} \left( \frac{1}{2\pi \overline{T}} \right)^{1/2} \frac{\Gamma\left(\frac{1}{4} - \frac{i \overline{\omega}}{2\pi \overline{T}} + i \mathcal{E}\right)}{\Gamma\left(\frac{3}{4} - \frac{i \overline{\omega}}{2\pi \overline{T}} + i \mathcal{E}\right)}
    \label{eq:sykConformalGreens}
\end{equation}
where $\theta$ and $\mathcal{E}$ characterize the spectral asymmetry and are related to the total charge $\mathcal{Q}$ by
\begin{equation}
    \begin{aligned}
        \mathcal{E} &= \frac{1}{2\pi} \ln \frac{\sin(\pi/4 + \theta)}{\sin (\pi/4 - \theta)}\,,
        \\
        \mathcal{Q} &= \frac{1}{2} - \frac{\theta}{\pi} - \frac{\sin(2\theta)}{4}\,.
    \end{aligned}
\end{equation}
We have the bounds $- \frac{\pi}{4} \leq \theta \leq \frac{\pi}{4}$ which implies $0 \leq \mathcal{Q} \leq 1$, and the particle-hole symmetric point is $\mathcal{Q} = \frac{1}{2}$.  Note that in contrast to the free fermion case, the SYK solution is most easily analyzed in the canonical ensemble with fixed charge $\mathcal{Q}$. 
These Green's functions satisfy the Schwinger-Dyson (SD) equation $\Sigma(\omega) G(\omega) = -1$, $\Sigma(\tau_1, \tau_2) =  J^2 G\left( \tau_1, \tau_2 \right)^{\frac{q}{2}} (-G(\tau_2, \tau_1))^{\frac{q}{2} - 1}$. We can evaluate the $\tau$ integrals of the top and bottom part of the Feynman diagram independently. Making use of the conformal SD equations, we find that each of these parts evaluates to
\begin{equation}
  \begin{aligned}
    &J^2 \int \dd{\tau_a} \dd{\tau_b} G(\tau_1, \tau_a) G(\tau_a, \tau_b)^{\frac{q}{2}} (-G(\tau_b, \tau_a))^{\frac{q}{2} - 1} G(\tau_b, \tau_2) 
    \\
    &= \int \dd{\tau_a} \dd{\tau_b} G(\tau_1, \tau_a) \Sigma(\tau_a, \tau_b) G(\tau_b, \tau_2) = - G(\tau_1, \tau_2)\,.
    \label{eq:sykIdentity}
  \end{aligned}
\end{equation}
A careful analysis of combinatoric factors from the disorder lines yields the result 
\begin{equation}
  \begin{aligned}
  \rho_o(\omega, \epsilon)  \frac{(q / 2)! (q / 2 - 1)!}{N^{q-1}} \Im \left[ G^R(\omega) \right] \Im \left[ G^R(\epsilon) \right] \,.
  \label{eq:sykrhoo}
  \end{aligned}
\end{equation}
For calculating $\rho_d$, the summation of ladder diagrams in Fig.~\ref{fig:complexSYKConnected} must be carried out. These ladder diagrams are well-studied in the SYK literature; in particular, evaluation in the strict conformal limit often leads to a divergent summation, with the regularizing near-conformal corrections taking a universal form that reflect the underlying dual quantum gravity description. This divergent summation is a consequence of ``resonant'' eigenfunctions of the ladder kernel which have eigenvalue unity.  Remarkably, there is no such effect in this class of ladder diagrams - because of the relation in Eq.~\ref{eq:sykIdentity}, we find that the conformal Green's function is an \textit{exact} eigenfunction with eigenvalue $q-1$, and therefore no resonance occurs. Because of this, the ladder diagrams may be evaluated via a geometric series to obtain the result
\begin{equation}
  \begin{aligned}
  \rho_d(\omega, \epsilon)  \frac{(q / 2)! (q / 2 - 1)!}{q^2 N^q} \Im \left[ G^R(\omega) \right] \Im \left[ G^R(\epsilon) \right] \,.
  \label{eq:sykrhod}
  \end{aligned}
\end{equation}

To leading order in $N^{-1}$, the conductance fluctuations are driven by $\rho_o$. The fact that $\rho_o$ factorizes into two copies of the spectral function leads to the simple result,
\begin{equation}
  \begin{aligned}
    \frac{\text{Var } \sigma}{\overline{\sigma}^2} = \frac{(q / 2)! (q / 2 - 1)!}{ N^{q-1}}\,.
    \label{eq:sykVariance}
  \end{aligned}
\end{equation}
The statement that the variance divided by the mean squared takes the above form holds for any linear statistic $A$ of the spectral function, $A = \int_{=\infty}^\infty \dd{\omega} A(\omega) \Im G^R(\omega)$.

\begin{figure}[htpb]
  \centering
  \includegraphics[width=\textwidth]{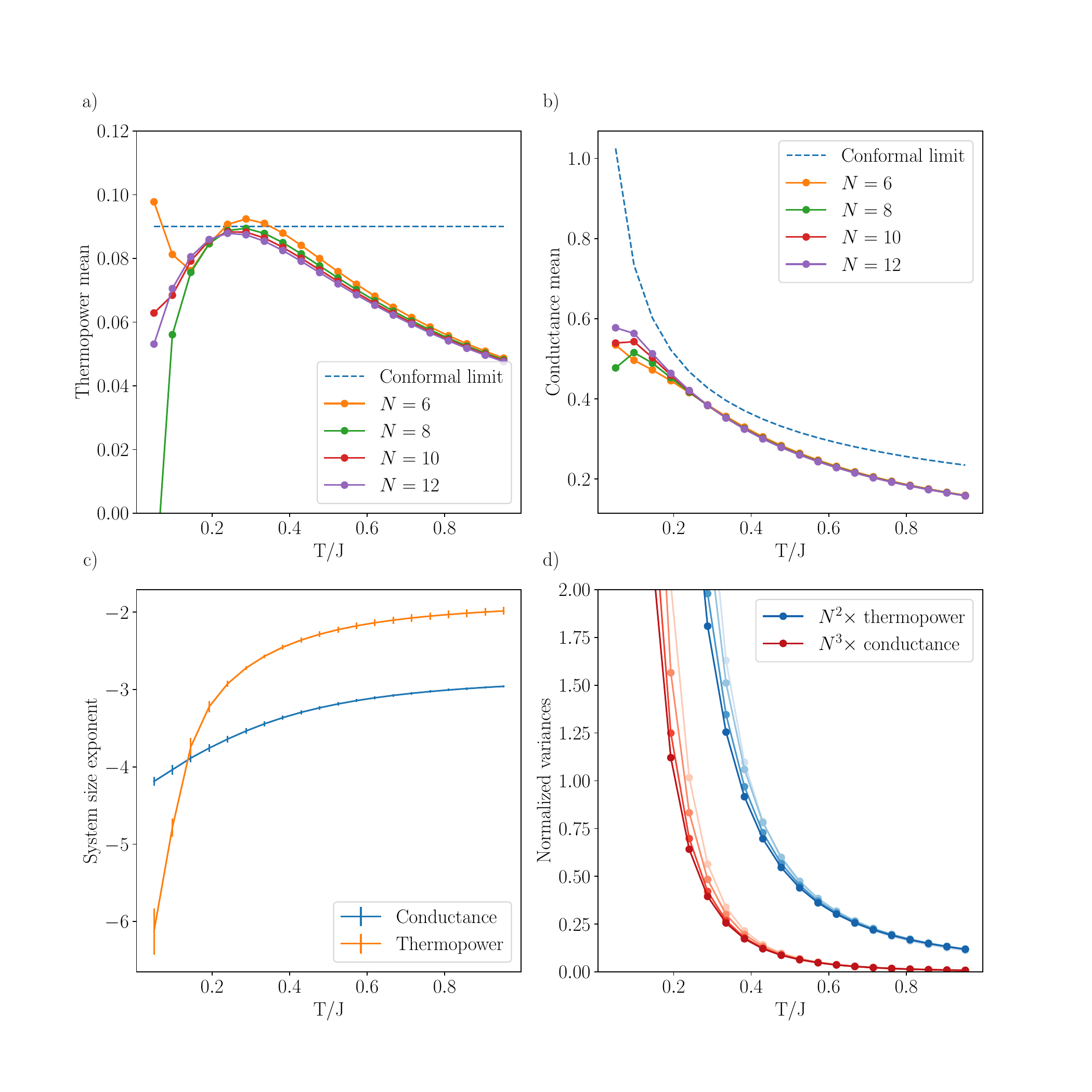}
  \caption{We present numerical results for the conductance and thermopower of a complex SYK model, for even system sizes $6 \leq N \leq 12$. All results are averaged over $10^5$ realizations with $J=1$, $\mu = 0.05$. a) The average thermopower, and conformal prediction. b) The average conductance, and conformal prediction. c) The system size scaling of the conductance and thermopower variance, obtained by fitting the variance as a function of $N$ to a power law at each temperature. d) We Temperature dependence of the normalized conductance and thermopower variance for $6 \leq N \leq 12$, both rescaled by their appropriate system size - $N^{-3}$ for conductance and $N^{-2}$ for thermopower. Darker plots indicate larger system sizes. }
  \label{fig:sykED}
\end{figure}

For the SYK model with four-fermion interactions, this gives
\begin{equation}
  \begin{aligned}
    \text{Var } \sigma = \left( \frac{e^2}{\hbar} \right)^2 \frac{1.04 \Gamma^2}{N^3 JT}\,.
  \end{aligned}
\end{equation}
We compare this result to numerical calculations of the conductivity variance using exact diagonalization, shown in Fig.~\ref{fig:sykED}. We also plot the mean values, which show decent agreement with their respective analytic predictions despite the relatively small system sizes.
Recall that Eq.~\ref{eq:sykVariance} is only valid in the conformal limit, where $\beta J \gg 1$. For finite size systems, we also require $\beta J \ll N$ due to Schwarzian fluctuations setting in at lower temperatures~\cite{kitaev2019, maldacena2016, gu2020, kobrin2021}. Exact diagonalization studies are restricted to small system sizes, with $N = 12$ the maximum size studied here. This implies a rather narrow temperature window where conformal behavior could be expected. We are unable to establish the constant temperature dependence predicted by Eq.~\ref{eq:sykVariance}; however, the predicted $N^{-3}$ scaling is expected to hold even at higher temperatures, away from the conformal limit, and this is validated by our numerical results. 

\subsection{Thermopower statistics}
We now analyze statistics of the thermopower, where the extensive entropy leads to a constant thermopower
\begin{equation}
  \begin{aligned}
  \overline{\Theta} = \frac{4\pi}{3 e} \mathcal{E}\,.
  \label{eq:sykThermopowerMean}
  \end{aligned}
\end{equation}
Recall that the thermopower is given by the ratio of two random variables, whose linear covariance vanished in the free fermion limit. Strikingly, the opposite behavior is true for an SYK model. To quantify this, we examine the Pearson correlation coefficient of the two random variables $A$ and $B$, 
\begin{equation}
  \begin{aligned}
    r_{A,B} \equiv \frac{\text{Cov}(A, B)}{\sqrt{\text{Var } A \times \text{Var } B}}  \,.
  \end{aligned}
\end{equation}
$r_{A, B}$ lies between $-1$ and $+1$ and measures the degree of correlation between two random variables. For the SYK model, a particular property of $\rho_o$ is that it that it factorizes into a product $\alpha(\omega) \alpha(\epsilon)$ to leading order in $N^{-1}$. Therefore, in sharp contrast to the Fermi liquid regime where the variables $\mathcal{L}_{11}$ and $\mathcal{L}_{12}$ were uncorrelated, we generically expect $r_{A, B} = 1 - \mathcal{O}(N^{-1})$. 

Despite this, we may still approximate our distribution as Gaussian. The approximation to normality of the distribution of two correlated Gaussian random variables follows along similar lines as the uncorrelated ratio~\cite{hinkley1969}, which we also discuss in more detail in Appendix~\ref{sec:ratioApp}. Defining $r$ as the correlation coefficient between $\mathcal{L}_{11}$ and $\mathcal{L}_{12}$, 
  \begin{equation}
  \begin{aligned}
    \frac{\text{Var } \Theta}{\overline{\Theta}^2} = \frac{\text{Var } \mathcal{L}_{11}}{\overline{\mathcal{L}_{11}}^2} + \frac{\text{Var } \mathcal{L}_{12}}{\overline{\mathcal{L}_{12}}^2} - \frac{2 r \sqrt{\text{Var } \mathcal{L}_{11} \times \text{Var } \mathcal{L}_{12}}}{\overline{\mathcal{L}_{11}} \overline{\mathcal{L}_{12}}}\,.
    \label{eq:correlatedVar}
  \end{aligned}
\end{equation}
Both $\mathcal{L}_{11}$ and $\mathcal{L}_{12}$ are linear statistics, so the conformal prediction of $ \frac{\text{Var } \mathcal{L}_{11}}{\overline{\mathcal{L}_{11}}^2} = \frac{\text{Var } \mathcal{L}_{12}}{\overline{\mathcal{L}_{12}}^2}$ leads to a vanishing thermopower variance. The leading order non-zero result in the conformal limit is hence suppressed by an additional factor of $N^{-1}$. However, high-temperature non-conformal corrections will still give an $\order{N^{1-q}}$ contribution.

Surprisingly, we find strong disagreement between this prediction and the exact diagonalization in Fig.~\ref{fig:sykED}. The thermopower variance in the temperature regime $\beta J \gg N^{-1}$ is well fit by a $N^{-2}$ scaling, rather than the $N^{-3}$ high-temperature contribution or the $N^{-4}$ conformal contribution. This arises due to an anomalous $N^{-2}$ scaling in the variance of the numerator, $\mathcal{L}_{12}$. As this quantity is proportional to the particle-hole asymmetry, we conjecture that this is related to additional fluctuations in the asymmetry not captured by our diagrammatic approach.
\section{Interplay between hoppings and interactions}
\label{sec:syk24}
In the previous sections, we have derived results for the conductance variance for both the limiting cases of non-interacting fermions with random hopping and a pure SYK model. In this section, we more carefully analyze the physically-relevant model with includes both random hopping and SYK terms.

Analysis of crossover behavior in these models has been performed previously~\cite{parcollet1999, song2017, kruchkov2020} for the average values of observables. The conclusion of these analyses is that there exists a coherence energy scale $E_{\text{coh}} \equiv \frac{t^2}{J}$ such that transport properties closely resemble the free fermion model for temperatures $T \ll E_{\text{coh}}$, with SYK behavior emerging for $T \gg E_{\text{coh}}$ (throughout this analysis, we assume $T \ll t\,, J$). The source of this behavior lies in the solution to the set of Schwinger-Dyson equations for the average value of the Green's function, which is exact in the large-$N$ limit:
\begin{equation}
  \begin{aligned}
    G(i\omega_n)^{-1} &= - i  \omega_n + \mu - t^2 G(i \omega_n)  - \Sigma(i \omega_n)\,,
    \\ 
    \Sigma(\tau)  &= - J^2 G^2(\tau) G(-\tau)\,.
    \label{eq:sdEqs}
  \end{aligned}
\end{equation}
It is this Green's function that displays a crossover at $T \sim E_{\text{coh}}$ from the free fermion-like solution to an SYK-like solution, which in turn leads to a crossover of the average values of transport properties. 

\begin{figure}[htpb]
  \centering
  \includegraphics[width=0.6\textwidth]{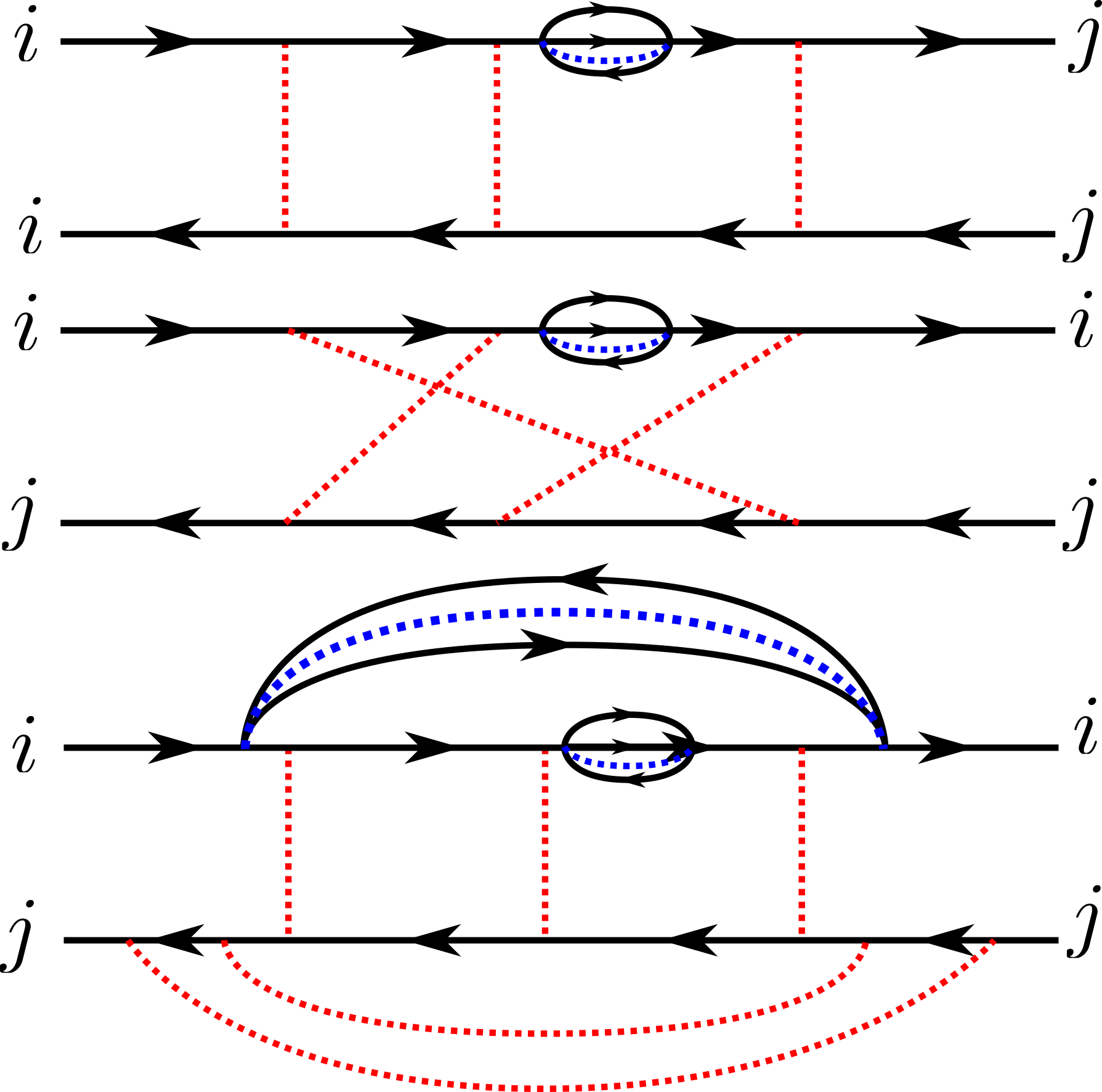}
  \caption{Ladder diagrams that contribute to the fluctuations of the single-particle spectral function to leading order in $N^{-1}$, for a model that includes both random single-particle hopping and SYK interactions. Disorder-averaging of the single-particle hopping (SYK interactions) is represented in red (blue). The structure of the diagrams are largely identical to the free fermion case illustrated in Fig.~\ref{fig:syk2ladders}, with the SYK interactions having the effect of renormalizing the average Green's functions. An exception to this is the additional set of diagrams, illustrated in the last diagram, which are qualitatively distinct from the free fermion limit. These diagrams only contribute to the diagonal covariance $\rho_d$ and hence will be neglected as they are suppressed by a factor of $N^{-1}$ relative to the off-diagonal covariances.}
  \label{fig:syk24ladders}
\end{figure}

In contrast, we claim that the variance of transport quantities displays a qualitatively different type of crossover behavior. This is a consequence of the free fermion variance in Sec~\ref{sec:freeFermion} and the SYK variance in Sec~\ref{sec:syk} containing different powers of $N$. Fluctuations driven by the randomness in SYK interactions are strongly suppressed relative to fluctuations driven by the random single-particle hopping. As a result, to leading order in $N^{-1}$, the free fermion Feynman diagrams in Fig.~\ref{fig:syk2ladders} - which exist for any arbitrarily small random hopping - are always the relevant ones for calculating fluctuation properties so long as the ratio $\frac{t}{J}$ does not scale with some inverse power of $N$. The effect of SYK interactions is to renormalize the average Green's functions, such that the Green's function that appear in Eq.~\ref{eq:syk2ladders} are given by the solution to Eq.~\ref{eq:sdEqs} rather than just the free fermion result. One can verify that to leading order in $N^{-1}$, the inclusion of SYK interactions does not modify the diagrammatic structure any further than this, with the exception of a class of diagrams illustrated in Fig.~\ref{fig:syk24ladders} - these diagrams only contribute to $\rho_d$ and hence will not be relevant for our analysis.

The key difference that results in the average values of thermoelectric properties being described by pure SYK for $T \gg E_\text{coh}$ and not their variances may be best understood conceptually within the framework of the $(G, \Sigma)$ action, which is worked out explicitly in Appendix~\ref{app:GSigma}. The intuition is as follows. For systems such as $H_{\text{dot}}$ with random all-to-all couplings, the fermionic degrees of freedom may be integrated out and the problem reformulated as a path integral over bilocal fields $G(\tau_1, \tau_2)$, $\Sigma(\tau_1, \tau_2)$, with an action that includes an explicit pre-factor of $N$; hence, the large-$N$ solution is described by the saddle point value of this action, which is precisely Eq.~\ref{eq:sdEqs}. The disorder-averaged spectral function, and in turn the average values of thermoelectric quantities, depend solely on this saddle-point solution. This is not true for fluctuations, which are subleading in $N^{-1}$ and is governed by replica off-diagonal fluctuations around the large-$N$ saddle point. The structure of the perturbation theory around the saddle point may be completely modified by the presence of a hopping term $t$ - Feynman diagrams proportional to $t$ may appear at lower orders in $N^{-1}$, and whose contributions will \textit{a priori} be dominant even in a parameter regime where the saddle point is well-described by the $t=0$ solution.

Our approach to studying the behavior of transport fluctuations for an interacting quantum dot will again involve calculating the single-particle covariance $\rho_{d, o}(\omega, \epsilon)$. We will work in the regime where $\omega, T \ll t, J$, and the average Green's function takes the universal form~\cite{parcollet1999}
\begin{equation}
  \begin{aligned}
  G(\omega, T) = \frac{1}{t} g\left( \frac{\omega}{E_{\rm coh}}, \frac{T}{E_{\rm coh}} \right) \equiv \frac{1}{t} g(\overline{\omega}, \overline{T}) \,,
  \end{aligned}
\end{equation}
where we define the dimensionless quantities $\overline{\omega} \equiv \omega / E_{\text{coh}}$, $\overline{T} \equiv T / E_{\text{coh}}$.
We find that the system sizes accessible to exact diagonalization are inadequate for establishing even the approximate crossover of the average Green's function; due to the narrow temperature window $N^{-1} \ll T \ll J\,, t$ where our analysis is valid, any crossover behavior is obscured by combination of high temperature or finite size effects. As a consequence, numerical results in this section will be restricted to self-consistent solutions of the Schwinger-Dyson equations given in Eq.~\ref{eq:sdEqs}.
\subsection{Fermi liquid regime}
\label{sec:tSYKFL}
For $\overline{T}, \overline{\omega} \ll 1$, it is known~\cite{parcollet1999} that $g^R(\overline{\omega}, \overline{T})$ has a Fermi liquid behavior. These properties can most simply stated at half filling ($\mu = 0$), where the Fermi liquid nature implies
\begin{equation}
  \begin{aligned}
  g^R(\overline{\omega} \ll 1, \overline{T} \ll 1) \approx -i \,.
  \label{eq:luttinger}
  \end{aligned}
\end{equation}
This behavior is determined by Luttinger's theorem, which for a generic charge $\mathcal{Q}$ says that
\begin{equation}
    \begin{aligned}
        \mu(\mathcal{Q}) - \Sigma(i 0^+) = \mu_0(\mathcal{Q}) 
    \end{aligned}
\end{equation}
where $\mu(\mathcal{Q})$ is the chemical potential necessary to tune to the charge $\mathcal{Q}$, and $\mu_0(\mathcal{Q})$ is that same value for the non-interacting ($J=0$) system. This fixes $G^R(\omega \rightarrow 0, T \rightarrow 0)$ to be that of the non-interacting Green's function, the latter of which we know has the property $\abs{g^R(\omega \rightarrow 0, T \rightarrow 0)}^2 = 1$ for generic filling. This property is sufficient for recovering the temperature-independent non-interacting prediction for the mean value of the conductance at low temperature, given in Eq.~\ref{eq:freeFermionConductanceMean} and likewise properly recovers the small $\omega, \epsilon$ divergence of $\rho_{d, o}$ given in Eq.~\ref{eq:divergences}. 
Although an explicit calculation of the conductance variance requires knowledge of the small frequency and temperature behavior of $g^R$, which is not fixed by Luttinger's theorem, the degree of the $T \rightarrow 0$ divergence and the assumption that small frequency/temperature corrections appear at linear order in $\overline{\omega}, \overline{T}$ imply from dimensional analysis that
\begin{equation}
  \begin{aligned}
    \text{Var } \sigma (T \ll E_{\text{coh}})  &\propto   \frac{1}{t^2 N \overline{T}} = \frac{1}{N J T}\,,
    \\
  \frac{\text{Var } \sigma (T \ll E_{\text{coh}}) }{\sigma (T \ll E_{\text{coh}})^2} &\propto   \frac{1}{N \overline{T}}\,.
  \end{aligned}
\end{equation}
This result is confirmed by calculating the conductance variance using the Green's function $G^R$ obtained from numerically solving the large-$N$ Schwinger-Dyson equations, shown in Fig.~\ref{fig:conductanceCrossover}.
\subsection{SYK Regime}
\label{sec:tSYK}
We now analyze the conductance fluctuations for $\overline{T} \gg 1$, where the average Green's function approaches the conformal SYK result given in Eq.~\ref{eq:sykConformalGreens}. The mean value of the conductance is then given by the pure SYK result in Eq.~\ref{eq:sykConductanceMean}. Using this form of the Green's function, we find that the $(\omega - \epsilon)^{-1}$ divergence of $\rho_o(\omega, \epsilon)$ is no longer present. The infinite sum of ladder diagrams that yields $\rho_o$ is convergent for large $\overline{T}$. Expanding in powers of $\overline{T}^{-1}$, we obtain the leading-order expression
\begin{equation}
  \begin{aligned}
    \text{Var } \sigma(T \gg E_{\text{coh}}) &= \left( \frac{e^2}{\hbar} \frac{\Gamma}{t \overline{T}} \right)^2 \frac{1}{N \pi^5}\frac{1}{2 \cos(2 \theta)} \left[\int_{-\infty}^\infty \dd{x}  \frac{e^x}{\left( 1+e^x \right)^2} \text{Im}\left[ h(x)^2 \right]  \right]^2\,,
    \\
    h(x) &\equiv e^{-i\theta}\frac{\Gamma\left( \frac{1}{4} - \frac{i x}{2\pi} + i \mathcal{E} \right) }{\Gamma\left( \frac{3}{4} - \frac{i x}{2\pi} + i \mathcal{E} \right) }\,.
    \label{eq:sykVarianceIntegral}
  \end{aligned}
\end{equation}
This integral must be done numerically; however, one can see that at the particle-hole symmetric point ($\theta = \mathcal{E} = 0$, $\mathcal{Q} = 1 / 2$), the integrand vanishes. We emphasize that this expression for the conductance variance is obtained by using the conformal SYK form of the Green's function \textit{and} taking to leading order a large-$\overline{T}$ expansion of the integral for the conductance variance, the latter of which is not a homogeneous function of $\overline{T}$. In particular, Eq.~\ref{eq:sykVarianceIntegral} does not imply that the conductance variance vanishes exactly in the conformal limit when $\mathcal{E} = 0$. Rather, the variance for $\mathcal{E} = 0$ is given by a subleading $\overline{T}^{-3}$ term. 
\begin{figure}[htpb]
  \centering
  \includegraphics[width=0.8\textwidth]{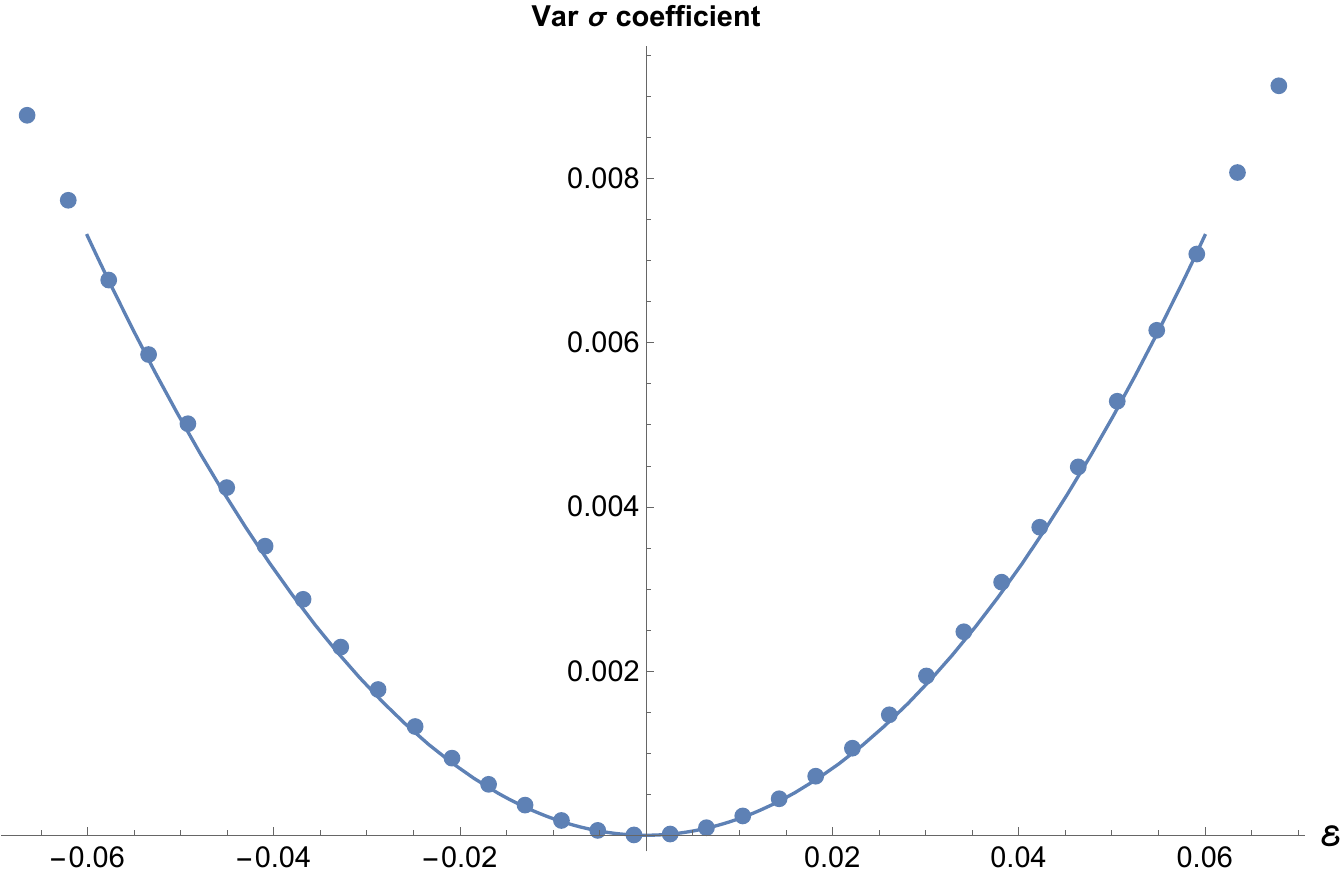}
  \caption{We plot the numerical coefficient of the leading-order conductance variance in the conformal SYK limit, obtained by a numerical evaluation of the integral in Eq.~\ref{eq:sykVarianceIntegral}, along with a quadratic approximation $4.05 \mathcal{E}^2$.}
  \label{fig:conformalCoeff}
\end{figure}
For general $\mathcal{Q}$, we find that the resulting expression is well-fit, see Fig.~\ref{fig:conformalCoeff} by the function
\begin{equation}
  \begin{aligned}
    \text{Var } \sigma(T \gg E_{\text{coh}}) &\approx \left( \frac{e^2}{\hbar} \frac{\Gamma}{t \overline{T}} \right)^2 \times  \frac{2.02 \mathcal{E}^2}{N}\,,
\\
\frac{\text{Var } \sigma (T \gg E_{\text{coh}})}{\sigma(T \gg E_{\text{coh}})^2} &\approx \frac{3.91 \mathcal{E}^2 }{N \overline{T}}\,.
  \end{aligned}
\end{equation}
We see that the conductance variance normalized by the mean squared has a $T^{-1}$ scaling, identical to the Fermi liquid regime. However, both quantities individually have distinct behavior, with the conductance variance scaling as $T^{-2}$ for $T \gg E_{\text{coh}}$ in contrast to the $T^{-1}$ scaling for $T \ll E_{\text{coh}}$. As an aside, we state the generalization to an $\text{SYK}_q$ model with $q$-fermion interactions; using the conformal Green's function gives a $\overline{T}^{\frac{8}{q} - 4}$ scaling of the conductance variance, and a $\overline{T}^{\frac{4}{q} - 2}$ scaling of the normalized conductance variance.

\begin{figure}[htpb]
  \centering
  \includegraphics[width=0.8\textwidth]{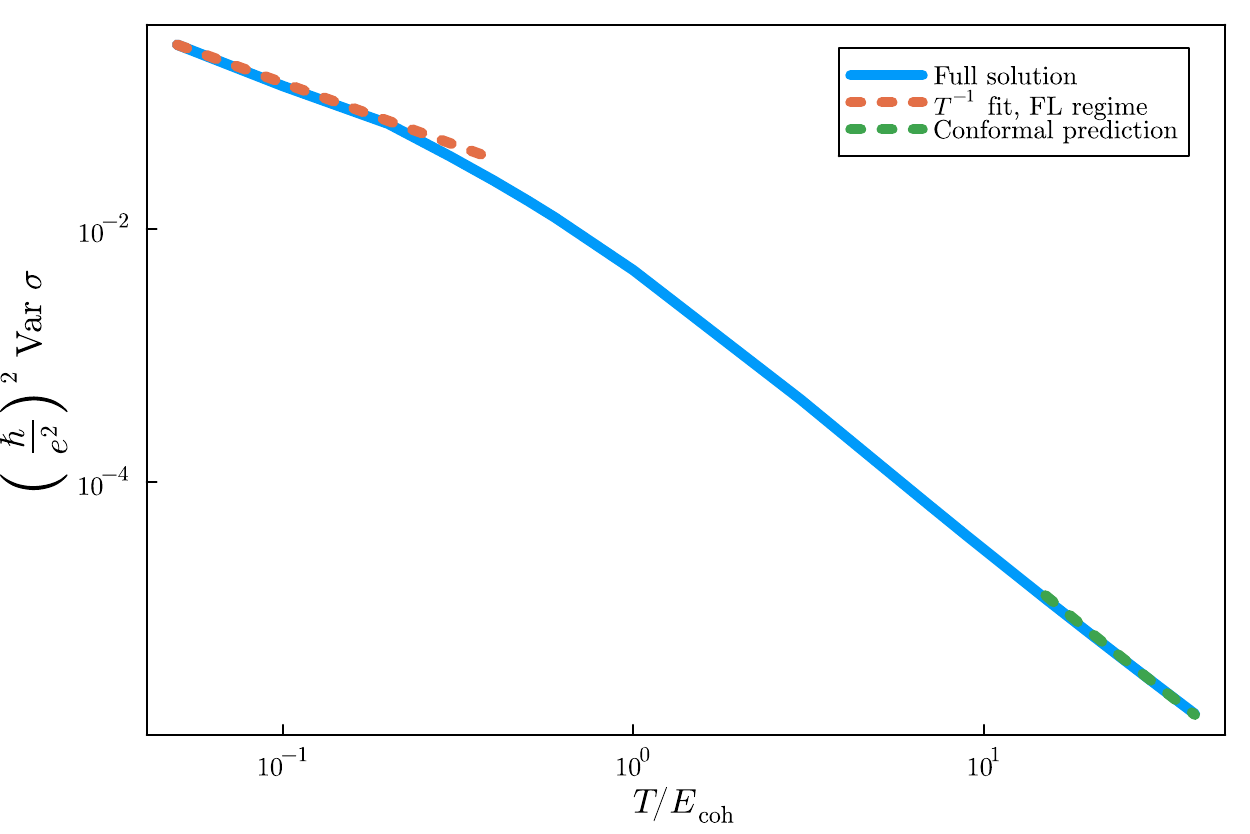}
  \caption{For parameters $J = 10$, $t = 0.1$, $\mathcal{Q} = 0.4$, $N=30$, and $\Gamma = 0.1$, we numerically solve for the leading order contribution to the conductance variance in the large-$N$ limit by solving the Schwinger-Dyson equations for the average Green's function over a range of temperatures. We demonstrate a crossover from $T^{-1}$ behavior at low temperatures, indicative of Fermi liquid behavior, to a more rapid $T^{-2}$ falloff at higher temperatures which reflects the average Green's function approaching the conformal SYK form.}
  \label{fig:conductanceCrossover}
\end{figure}

This crossover behavior is demonstrated in Fig.~\ref{fig:conductanceCrossover}, where we solve for the conductance variance given the form of the Green's function covariance in Eq.~\ref{eq:syk2ladders}, where we use the average Green's function $G^R(\omega)$ obtained from a full self-consistent solution of the Schwinger-Dyson equations in real time. Details on the numerical implementation for solving the real-time Schwinger-Dyson equations can be found in~\cite{song2017}. We note a unique difficulty in calculating the conductance variance not present in the average value, which comes from the denominator $1 - t^2 G^R(\omega) G^A(\omega)$ in the Green's function covariance. As discussed previously, it is characteristic of a Fermi liquid that this denominator goes to zero as $T \rightarrow 0$. As a consequence, the accuracy with which one must numerically solve for $G^R(\omega)$ diverges as $T \rightarrow 0$; small errors at low temperatures can easily lead to an unphysical divergence in the conductance variance. Our self-consistent solution for $G^R(\omega)$ utilizes a grid of $2^{28}$ frequency points on the real axis, which gives a sufficiently accurate solution down to $T / E_{\text{coh}} \approx 0.03$ and is enough to recover the predicted $T^{-1}$ scaling at low temperatures.   
\subsection{Thermopower statistics}
The mean thermopower in a model with both random hopping and SYK interactions displays a crossover from the linear temperature scaling characteristic of a Fermi liquid for $T \ll E_{\text{coh}}$ to the constant SYK value for $T \gg E_{\text{coh}}$. The coefficient of the mean thermopower in the Fermi liquid regime receives a renormalization due to the presence of SYK interactions, from $\overline{\Theta} \sim (et)^{-1} T$ in the free fermion model to $\overline{\Theta} \sim (e E_\text{coh})^{-1} T$. This is not true for the mean conductance, whose value for $T\rightarrow 0$ is determined by the zero-frequency spectral density and is fixed by Luttinger's theorem, Eq.~\ref{eq:luttinger}.

We now discuss the crossover behavior of the thermopower variance. For $T \ll E_{\text{coh}}$, the thermopower variance follows from the free fermion analysis in Section~\ref{sec:freeFermion} and diverges as $\overline{T}^{-1}$ for low temperatures, albeit with a renormalized coefficient. For $T \gg E_{\text{coh}}$, we find that the Pearson correlation coefficient $r$ between $\mathcal{L}_{11}$ and $\mathcal{L}_{12}$ is $1$ to leading order in $\overline{T}$. We apply Eq.~\ref{eq:correlatedVar}, which gives the thermopower variance in terms of $r$ and the statistics of $\mathcal{L}_{11}$ and $\mathcal{L}_{12}$, where now we have
\begin{equation}
  \begin{aligned}
    \text{Var } \mathcal{L}_{11} &=   \left( \frac{\Gamma}{\hbar} \right)^2 \times \frac{2.02 \mathcal{E}^2}{N t^2 \overline{T}^2} \,, \quad \overline{\mathcal{L}_{11}} = \frac{\Gamma}{\hbar} \frac{0.72}{t\sqrt{\overline{T}}} \,,\\
    \text{Var } \mathcal{L}_{12} &= \left( \frac{\Gamma}{\hbar} \right)^2 \times \frac{0.07}{Nt^2} \,, \quad \overline{\mathcal{L}_{12}} = \frac{\Gamma}{\hbar}\frac{3.01 \overline{T}^{1/2}\mathcal{E}}{t}\,. \\
    \label{eq:coeffVars}
  \end{aligned}
\end{equation}
All of the terms in Eq.~\ref{eq:correlatedVar} decay as $\overline{T}^{-1}$, which implies that in the limit $r=1$,
\begin{equation}
    \begin{aligned}
        \frac{\text{Var } \Theta}{\overline{\Theta}^2}  = \frac{1}{N \overline{T}} \left(1.97\abs{\mathcal{E}} - 0.09 \abs{\mathcal{E}}^{-1} \right)^2\,.
    \end{aligned}
\end{equation}
The coefficient is rather striking, as it predicts a suppression of this leading-order variance at a critical value of the particle-hole asymmetry $\abs{\mathcal{E}_c} \approx 0.24$. Recall that this leading-order suppression happens generically for a pure SYK model - this is a consequence of expanding around the limit of perfect correlation between $\mathcal{L}_{11}$ and $\mathcal{L}_{12}$, along with the identity $\frac{\text{Var} \mathcal{L}_{11}}{\overline{\mathcal{L}_{11}}^2} = \frac{\text{Var} \mathcal{L}_{12}}{\overline{\mathcal{L}_{12}}^2}$. The latter identity is not true generically in this model, but only occurs at the aforementioned fine-tuned value $\mathcal{E}_c$. This value of $\mathcal{E}$ corresponds to a rather large particle-hole asymmetry however, $\mathcal{Q}_c \approx \frac{1}{2} \pm 0.41$, and is hence not easily accessible.

\section{Conclusion}
\label{sec:conc}

We have analyzed the fluctuations of thermoelectric transport properties in strongly-correlated quantum dots. Despite the apparent simplicity of our microscopic model due to its exact large-$N$ solution, this saddle point only describes the mean value of transport quantities; higher-order moments are controlled by replica off-diagonal fluctuations around this saddle point, and as such require a more unconventional analysis. We find distinct system size scalings for these fluctuations in a free fermion model $(N^{-1})$ and an SYK model $(N^{-3})$. The SYK prediction is qualitatively changed by the inclusion of a small random hopping, which we find is able to drive conductance fluctuations at the same order as the free fermion prediction. However, we still find distinct temperature scalings, with a $T^{-2}$ suppression for temperatures above the coherence energy in contrast to the $T^{-1}$ scaling at lower temperatures predicted by the free fermion result. 

Our main analytic results for the conductance, $\sigma$ were summarized in Section~\ref{sec:intro}. We also computed the thermopower, $\Theta$. The mean thermopower vanishes linearly with $T$ in the Fermi liquid regime (see Eq.~\ref{Thetalinear}), while the SYK regime has a $T$-independent thermopower (see Eq.~\ref{eq:sykThermopowerMean}). Furthermore, the finite $N$ Schwarzian corrections are quite small for the mean thermopower in the SYK regime \cite{kruchkov2020}. These features make the thermopower an ideal probe for detecting the SYK regime in experiments. However, analytic computations of the sample-to-sample fluctuations in the thermopower are not straightforward because the expression for the thermopower involves the ratio of electron Green's functions. We made partial analytic progress assuming small Gaussian fluctuations about the mean of both the numerator and the denominator, and also obtained numerical exact-diagonalization results for small values of $N$. Our main results are as follows. For a free fermion model, the thermopower variance scales as $t \left(N T\right)^{-1}$, in good agreement with numerical results. For a pure SYK model, we find surprisingly that the leading order $N^{-3}$ contribution to the thermopower variance vanishes in the conformal limit ($T \ll J$) due to perfect correlation between the numerator and denominator. Fluctuations in this regime are hence governed by a combination of high-temperature and $\order{N^{-4}}$ corrections, although we are unable to verify this behavior numerically due to anomalous $\order{N^{-2}}$ fluctuations. For a model with both random hopping and SYK interactions, our predictions once again are qualitatively modified. The scaling of the variance in the low temperature Fermi liquid regime is suppressed from the free fermion result $t \left( NT \right)^{-1}$ by an additional factor of $t / J$. In the SYK regime, the scaling is identical, albeit arising from distinct mechanisms. A noteworthy feature in the SYK regime is that this leading-order variance vanishes at a critical value of the particle-hole asymmetry $\mathcal{E}_c$, in which case the first non-zero contribution scales as $N^{-1} (T / E_{\text{coh}})^{-2}$.

A more careful treatment of the effects of the coupling between the quantum dot and the leads may reveal richer physics. In this work, we restrict our parameter regime to a ``closed'' quantum dot, where the coupling to the leads is the smallest energy scale in the system and transport quantities follow from the properties of the isolated quantum dot. A more robust framework for treating the effects of the leads can be developed by treating both the single-particle hopping in the leads and the coupling to the quantum dot as random variables, for which an exact (in the large-$N$ limit) set of Schwinger-Dyson equations can be obtained for the non-equilibrium Green's functions~\cite{can2019}. The mean value of the conductance has been studied using this framework, although the effects of single-particle hopping within the quantum dot were not considered. In addition to treating conductance fluctuations within this framework, an analysis of the effects of inter-dot single-particle hopping, which was not considered in~\cite{can2019}, may lead to new predictions even in the average value of transport properties.

The nature of conductance fluctuations for a pure SYK model is also deserving of further analysis. The results we present are confined to the conformal regime. Deviations from this prediction at higher temperatures can be captured by an analysis of the large-$N$ numerical solution to the Schwinger-Dyson equations, and low-temperature deviations may be understood analytically through Schwarzian fluctuations. This analysis is also expected to give greater agreement with numerical results for small system sizes, where clear agreement with the conformal prediction is absent.

\subsection*{Acknowledgements}

We thank Alex Kruchkov for significant discussions at the initial stages of this work. We also thank Yigal Meir for helpful comments. This research was supported by the U.S. National Science Foundation grant No. DMR-2245246 and by the Simons Collaboration on Ultra-Quantum Matter which is a grant from the Simons Foundation (651440, S.S.). PK and LA acknowledge support from ONR MURI (N00014-21-1-2537).

\appendix
\section{Path integral calculation of fluctuations}
\label{app:replica}
In this Appendix, we review the procedure for calculating the fluctuations of observables in disordered systems using the path integral approach. 

Calculating statistical quantities in disordered systems, such as averages and variances, is in general a non-trivial task. This arises from the fact that correlation functions such as $G(\tau - \tau')$ for a given disorder realization $J_{i j k l}$ (this notation is specific to an SYK model, which we will use without loss of generality) are given by functional integrals of the form
\begin{equation}
  \begin{aligned}
G(\tau - \tau') = \frac{1}{N}   \frac{\int \mathcal{D} c^\dagger \mathcal{D}c \, \sum_i c_i^\dagger(\tau) c_i(\tau') e^{-S[c\,, c^\dagger \,, J_{i j k l}]}}{\int \mathcal{D} c^\dagger \mathcal{D}c \,  e^{-S[c\,, c^\dagger \,, J_{i j k l}]}}\,.
    \label{eq:greens}
  \end{aligned}
\end{equation}
The mean of this quantity over an ensemble $P(J_{i j k l})$ is given by integrating it over all realizations of $J_{i j k l}$. This averaging cannot simply be done, as Eq.~\ref{eq:greens} is a ratio of two quantities. What can be done analytically is carry out the average of the numerator and denominator separately - this constitutes treating the random variables $J_{i j k l}$ on the same footing as our physical variables $c_i^\dagger\,, c_i$. Treating the disorder average properly requires techniques such as the replica trick~\cite{edwards1975}, which we will employ here. Supersymmetric techniques have also been developed for dealing with these averages~\cite{efetov1983}, which is the primary method used for calculating conductance fluctuations of free electrons and generally yields more reliable results than the replica approach, the latter of which requires a generally-uncontrolled analytical continuation of the number of replicas $M \rightarrow 0$. However, these supersymmetric techniques are not appropriate for including the effects of strong interactions. Recent advances have generalized these supersymmetry techniques to a particular variant of the SYK model~\cite{sedrakyan2020}, and an interesting direction for future research would be to see whether such an approach is applicable to our model or a variant thereof that would allow for more controlled calculations of transport fluctuations.

Here, we make explicit the setup we use to calculate fluctuations of quantities like $G(i\omega)$. What we are interested in is the covariance of the Green's function at different frequencies, such as $\frac{1}{N^2} \sum_{ij} \left[\overline{G_{ii}(i\omega) G_{jj}(i\epsilon)} - \overline{G_{ii}(i\omega)}\,\overline{G_{jj}(i\epsilon)} \right]$. Using the replica trick, we can rewrite the product of Green's functions $G(\tau_1 - \tau_2) G(\tau_3 - \tau_4)$ as a functional integral taken over two copies of fermionic variables,$c_{i}^a\,, c_{i}^{\dagger a}\,, \tilde{c}_{i}^{a'}\,, \tilde{c}_{i}^{\dagger a'}$, with $i$ a site index and $a\,, a'$ replica indices,
\begin{equation}
  \begin{aligned}
    \lim_{M\,, M' \rightarrow 0}\frac{1}{N^2 M M'}\sum_{\substack{1 < a < M\\ 1 < a' < M'}} \int  \sum_{i, j} c_{i}^{\dagger a}(\tau_1) c_{i}^a(\tau_2) \tilde{c}_{j}^{\dagger a'}(\tau_3) \tilde{c}_{j}^{a'}(\tau_4) e^{-\sum_a S[c_{i}^{\dagger a} \,, c_{i}^a, J_{ij kl}] -\sum_{a'} S[\tilde{c}_{i}^{\dagger a'} \,, \tilde{c}_{i}^{a'}, J_{ij kl}] }
    \label{eq:replicas}
  \end{aligned}
\end{equation}
We can dispense of the independent replica summations and the distinction between $c$ and $\tilde{c}$ by combining them into an enlarged summation,
\begin{equation}
  \begin{aligned}
    \lim_{M \rightarrow 0}\frac{1}{N^2 M^2}\sum_{\substack{1 < a\,, b < M\\ a \neq b}} \int  \sum_{i, j} c_{i}^{\dagger a}(\tau_1) c_{i}^a(\tau_2) c_{j}^{\dagger b}(\tau_3) c_{j}^{b}(\tau_4) e^{-\sum_d S[c_{i}^{\dagger d} \,, c_{i}^d, J_{ij kl}]  }
    \label{eq:replicas2}
  \end{aligned}
\end{equation}
The action $S$ is a function of the random variables $J_{i j k l}$, and the disorder average is performed over the above quantity. Doing this induces interactions between the different replicas. Subtracting off the disconnected contribution, $\overline{G(\tau_1 - \tau_2)} \, \overline{G(\tau_3 - \tau_{4})}$ constitutes disregarding contributions that do not contain any interactions between the two replica indices. An analogous treatment of the off-diagonal covariance, $\frac{1}{N^2} \sum_{ij} \left[\overline{G_{ij}(i\omega) G_{ji}(i\epsilon)} - \overline{G_{ij}(i\omega)}\,\overline{G_{ji}(i\epsilon)} \right]$ leads to an expectation value of the form $c_i^{\dagger a}(\tau_1) c_i^b(\tau_2) c_j^{\dagger b}(\tau_3) c_j^a(\tau_4)$.

For our calculations, we will proceed perturbatively starting from the replica-symmetric saddle point. If we use this as our starting point, our propagators will remain replica-symmetric to all orders in perturbation theory~\cite{arefeva2019}. It has been shown that for free fermions, this approximation is sufficient for accurately recovering the leading-order contribution to the mean value of $G(\tau_1 - \tau_2)$, although $N^{-1}$ corrections require replica-off-diagonal saddles~\cite{kamenev1999}. For four-point functions like Eq.~\ref{eq:replicas}, it is known that a replica-diagonal ansatz is insufficient for reproducing the full spectral correlations of random matrix theory~\cite{verbaarschot1985} for small $\order{N^{-1}}$ energy differences, but can be recovered by considering off-diagonal saddle manifolds~\cite{kamenev1999}. This discrepancy is not relevant for our analysis, as we will only be interested in spectral correlations over $\order{T}$ energy differences. 
\section{Replica off-diagonal fluctuations in the \texorpdfstring{$(G, \Sigma)$}{GSigma} action}
\label{app:GSigma}
The calculation of the Green's function covariances may be performed within the formalism of the $(G, \Sigma)$ path integral, which we describe here. Although this perspective does not provide a direct computational advantage over the fermionic diagram approach in the main text - all non-trivial integrals are still present - it admits an explicit $N^{-1}$ expansion, in contrast with the diagrammatic approach in the main text where the task of writing down all diagrams that contribute at a given order requires careful analysis of index summations. The approach here is more easily generalizable to the calculation of higher order moments, and also provides a more general framework for understanding which observables obey a straightforward crossover from SYK-like to Fermi liquid-like as a function of temperature and which ones have more subtle crossover behavior - the former are functions of only the saddle point solutions of the $(G, \Sigma)$ path integral, whereas the latter are properties of fluctuations around the saddle point. Here, we rederive the off-diagional Green's function covariance, $\rho_o$, using this formulation.

We begin with a derivation of the $(G, \Sigma)$ path integral. Recall that our Hamiltonian is given by
\begin{equation}
  \begin{aligned}
    H &= \frac{1}{(2N)^{3 / 2}}\sum_{ij; kl = 1}^N J_{ij ; kl} c_i^\dagger c_j^\dagger c_k c_l
  + \frac{1}{N^{1 / 2}}\sum_{ij = 1}^N t_{ij} c_i^\dagger c_j - \mu \sum_i c_i^\dagger c_i
  \end{aligned}
\end{equation}
where $J_{ij; kl}$ and $t_{ij}$ are complex random numbers with zero mean and variances $J^2$ and $t^2$, respectively.
In path integral form, we have the partition function
\begin{equation}
  \begin{aligned}
    \overline{Z[h]^M} &= \int \mathcal{D} J \mathcal{D} t \mathcal{D}c \mathcal{D}c^\dagger e^{-\sum_{a=1}^M S_a[J] }
    \\
    S_a[J] &= \sum_{ij}\int \dd{\tau} c_i^{\dagger a}(\tau)\left[ \left(\partial_\tau - \mu \right) \delta_{ij} + \frac{t_{ij}}{N^{1 / 2}}\right] c_j^a(\tau) 
    \\
    &+ \frac{1}{(2 N)^{3 / 2}} \sum_{ij; kl} \int \dd{\tau} J_{ij;kl} c_{i}^{\dagger a}(\tau)  c_{j}^{\dagger a}(\tau)  c^a_{k}(\tau)  c^a_{l}(\tau)  \\
  \end{aligned}
\end{equation}

Integrating over disorder, our path integral becomes
\begin{equation}
  \begin{aligned}
    Z[h] &= \int \mathcal{D}c \mathcal{D} c^\dagger  e^{-S} \\
    S &= \sum_{a, i} \int \dd{\tau} c_i^{\dagger a}( \partial_\tau - \mu) c_i^a - \sum_{a\,, b}\int \dd{\tau_1} \dd{\tau_2}\Bigg[ \frac{N J^2}{4}\left( \frac{1}{N}\sum_{i} c_i^{\dagger a}(\tau_1) c_i^b(\tau_2) \right)^{2}\left( \frac{1}{N}\sum_{i} c_i^{\dagger b}(\tau_2) c_i^a(\tau_1) \right)^{2} 
    \\
  &- \frac{N t^2}{2} \left( \frac{1}{N}\sum_{i} c_i^{\dagger a}(\tau_1) c_i^b(\tau_2) \right)\left( \frac{1}{N}\sum_{i} c_i^{\dagger b}(\tau_2) c_i^a(\tau_1) \right)  \Bigg]  \\
  \end{aligned}
\end{equation}
We now insert the field 
\begin{equation}
  \begin{aligned}
    G^{ab}(\tau_1\,, \tau_2) \equiv \frac{1}{N} \sum_i c^{\dagger a}_i(\tau_1)c_i^b(\tau_2)
  \end{aligned}
\end{equation}
where the equivalence is enforced with a Lagrange multiplier $\Sigma^{ab}(\tau_1, \tau_2)$. The $c\,, c^\dagger$ fields can be integrated out to yield the action
\begin{equation}
  \begin{aligned}
    \frac{S[G\,, \Sigma\,, h]}{N} &= - \ln \det (- \partial_\tau + \mu - \Sigma) - \sum_{a\,, b} \int \dd{\tau_{1, 2}} \Bigg( \Sigma^{ab}(\tau_1, \tau_2) G^{ba}(\tau_2, \tau_1) 
      \\
      &+ \frac{J^2}{4} \left(G^{ab}(\tau_1, \tau_2) G^{ba}(\tau_2, \tau_1)\right)^{2}  - \frac{t^2}{2} G^{ab}(\tau_1, \tau_2) G^{ba}(\tau_2, \tau_1) \Bigg)   \,.
    \label{eq:fullGSigma}
  \end{aligned}
\end{equation}

We take the replica-diagonal saddle point, $G^{ab}(\tau_1, \tau_2) = \delta_{ab} G(\tau_1 - \tau_2)$ and likewise for $\Sigma^{ab}$. The replica-diagonal Schwinger-Dyson equations are given by Eq.~\ref{eq:sdEqs} in the main text - as emphasized earlier, it is the solution to this set of equations that displays a crossover from SYK-like for $T \gg E_{\text{coh}}$ to Fermi liquid-like for $T \ll E_{\text{coh}}$. This saddle-point solution does not contribute to the Green's function covaraince; to obtain a non-zero value, we must consider fluctuations around it, $G^{ab}(\tau_1, \tau_2) \equiv \delta_{ab}G(\tau_1 - \tau_2) +\delta G^{ab}(\tau_1, \tau_2)$. 

In this representation, our observables of interest are
\begin{equation}
  \begin{aligned}
    g_o(\tau_{1, 2, 3, 4}) &\equiv \frac{1}{N^2}\sum_{ij} \left[\overline{G_{ij}(\tau_1 - \tau_2) G_{ji}(\tau_3 - \tau_4)} - \overline{G_{ij}(\tau_1 - \tau_2)} \, \overline{ G_{ji}(\tau_3 - \tau_4)}\right]
    \\
    &= \langle G^{ab}(\tau_1 - \tau_2) G^{ba}(\tau_3 - \tau_4) \rangle - \frac{1}{N} \langle G^{aa} (\tau_1 - \tau_2) G^{bb} (\tau_3 - \tau_4) \rangle \\
  \end{aligned}
\end{equation}
for $a \neq b$. Note the subleading correction in $g_o$, which arises from the $i = j $ term in the disconnected contribution (the ``standard'' disconnected part of $g_o$ vanishes due to the fact that $\langle G^{ab} \rangle = 0$ for fluctuations around the replica-diagonal saddle point). 

\begin{figure}[htpb]
  \centering
  \includegraphics[width=0.5\textwidth]{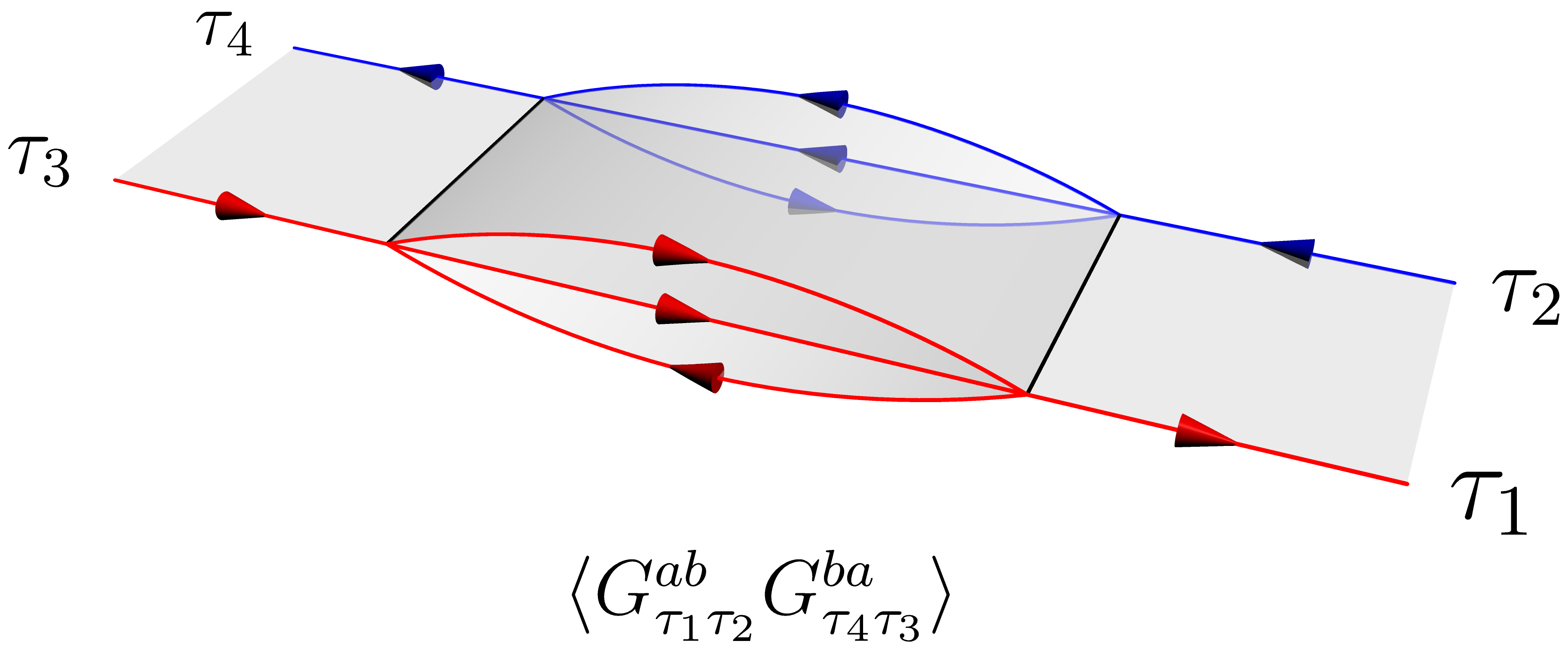}
  \caption{We illustrate Feynman diagrams that contribute to the off-diagonal Green's function covariance in a pure SYK model. For a model that includes random hoppings, there exists a non-trivial contribution in the bare $\delta G^{ab} \delta G^{ba}$ propagator; for a pure SYK model, this contribution is subtracted off exactly in the covariance and one must include the leading order correction to obtain a non-zero result.}
  \label{fig:oDiagrams}
\end{figure}

These replica off-diagonal observables vanish at the replica-diagonal saddle point. To find the leading order non-zero result, we expand the action around its saddle-point solution. The expansion of everything other than the determinant is rather straightforward. For evaluation of the determinant, we use Jacobi's formula
\begin{equation}
  \begin{aligned}
    &\frac{1}{\det(-\partial_\tau + \mu - \Sigma)} \frac{\partial \det (-\partial_\tau + \mu - \Sigma)}{\partial \Sigma^{ab}(\tau_1, \tau_2)} =  -\Tr \left[ \left( - \partial_\tau + \mu - \Sigma\right)^{-1} \frac{\partial \Sigma}{\partial \Sigma^{ab}(\tau_1, \tau_2)}   \right] 
    \\
    &= -\left[\left( -\partial_\tau + \mu - \Sigma \right)^{-1} \right]^{ba}\left( \tau_2, \tau_1 \right) =- \delta_{ab}G(\tau_2 - \tau_1)   \\
  \end{aligned}
\end{equation}
where in the final line we evaluate the expression at the replica-diagonal saddle point. 
To second order, we use
\begin{equation}
  \begin{aligned}
    &\frac{1}{\det(-\partial_\tau + \mu - \Sigma)} \frac{\partial^2 \det (-\partial_\tau  + \mu- \Sigma)}{\partial \Sigma^{ab}(\tau_1, \tau_2) \partial \Sigma^{cd}(\tau_3, \tau_4)} 
    \\
    &= -\frac{1}{\det(-\partial_\tau + \mu - \Sigma)} \pdv{}{\Sigma^{cd}(\tau_3, \tau_4)} \left[ \det(-\partial_\tau + \mu - \Sigma) \Tr \left[ \left( - \partial_\tau + \mu - \Sigma\right)^{-1} \frac{\partial \Sigma}{\partial \Sigma^{ab}(\tau_1, \tau_2)}   \right] \right]
    \\
    &= \delta_{ab} \delta_{cd} G(\tau_2 - \tau_1) G(\tau_4 - \tau_3) - \Tr\left[ \delta \Sigma^{ab} G \delta \Sigma^{ba} G \right] \\
  \end{aligned}
\end{equation}
This leads to the quadratic action
\begin{equation}
  \begin{aligned}
    &\frac{\delta S\left[ \delta G\,, \delta \Sigma\right]}{N} = \sum_{ab} \Big[ \frac{1}{2} \Tr\left[ G \delta\Sigma^{ab} G \delta \Sigma^{ba} \right] - \int \dd{\tau_1} \dd{\tau_2} \delta G^{ab}(\tau_1, \tau_2) \left[\delta \Sigma^{ba}(\tau_2, \tau_1)  - \frac{t^2}{2} \delta G^{ba}(\tau_2, \tau_1) \right]
    \\
  &- \frac{J^2  \delta_{ab}}{2} \int \dd{\tau_1} \dd{\tau_2} \Bigg(2 G(\tau_1, \tau_2) G(\tau_2, \tau_1) \delta G^{aa}(\tau_1, \tau_2) \delta G^{aa}(\tau_2, \tau_1) 
  \\
&+ G(\tau_1, \tau_2)^{2}  \delta G^{aa}(\tau_2, \tau_1) \delta G^{aa}(\tau_1, \tau_2)\Bigg) 
  \end{aligned}
\end{equation}
\begin{figure}[htpb]
  \centering
  \includegraphics[width=0.8\textwidth]{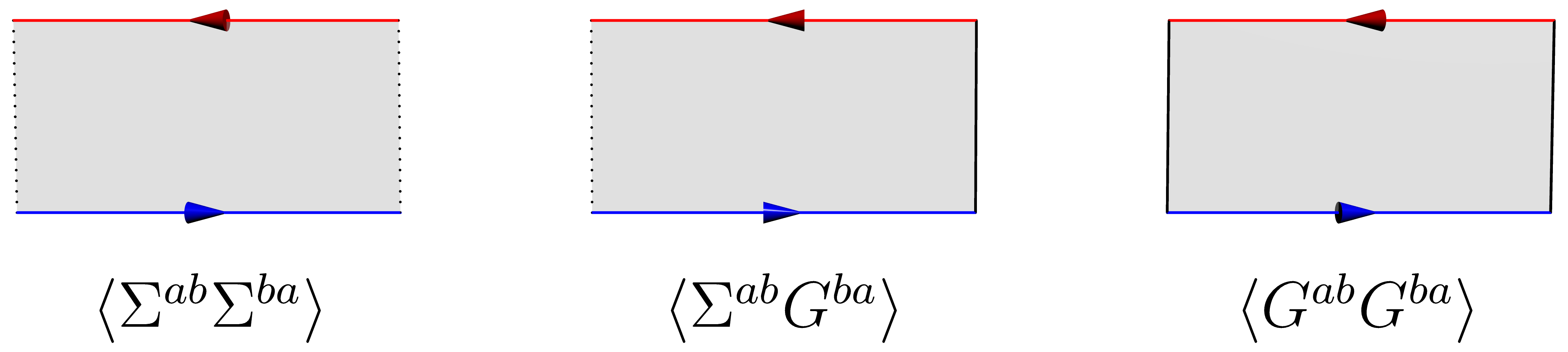}
  \caption{We illustrate the propagators for use in a diagrammatic expansion in $N^{-1}$ around the saddle point of the $(G\,, \Sigma)$ action. The fields $G$ and $\Sigma$ are a function of two times and two replica indices, which necessitates the sheet-like representation above. The colors indicate different replica indices $a\,, b$, and solid (dotted) lines indicate a $G$ ($\Sigma$) field. }
  \label{fig:feynmanProps}
\end{figure}
\begin{figure}[htpb]
  \centering
  \includegraphics[width=0.8\textwidth]{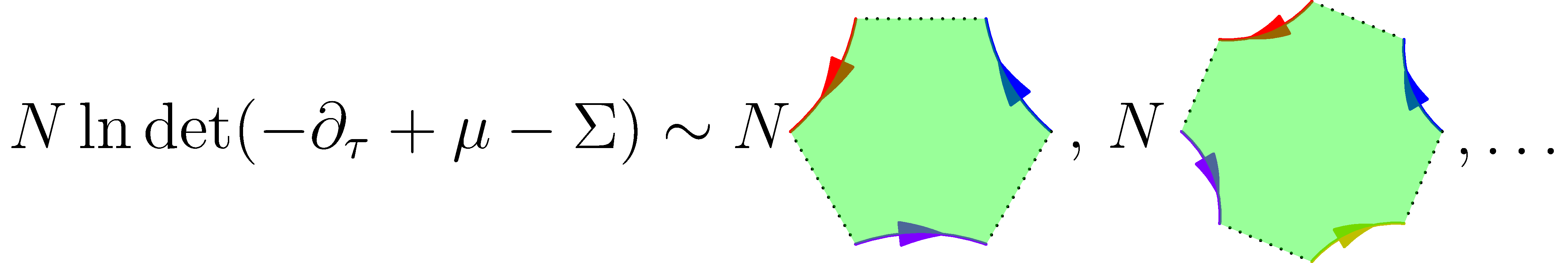}
  \includegraphics[width=0.4\textwidth]{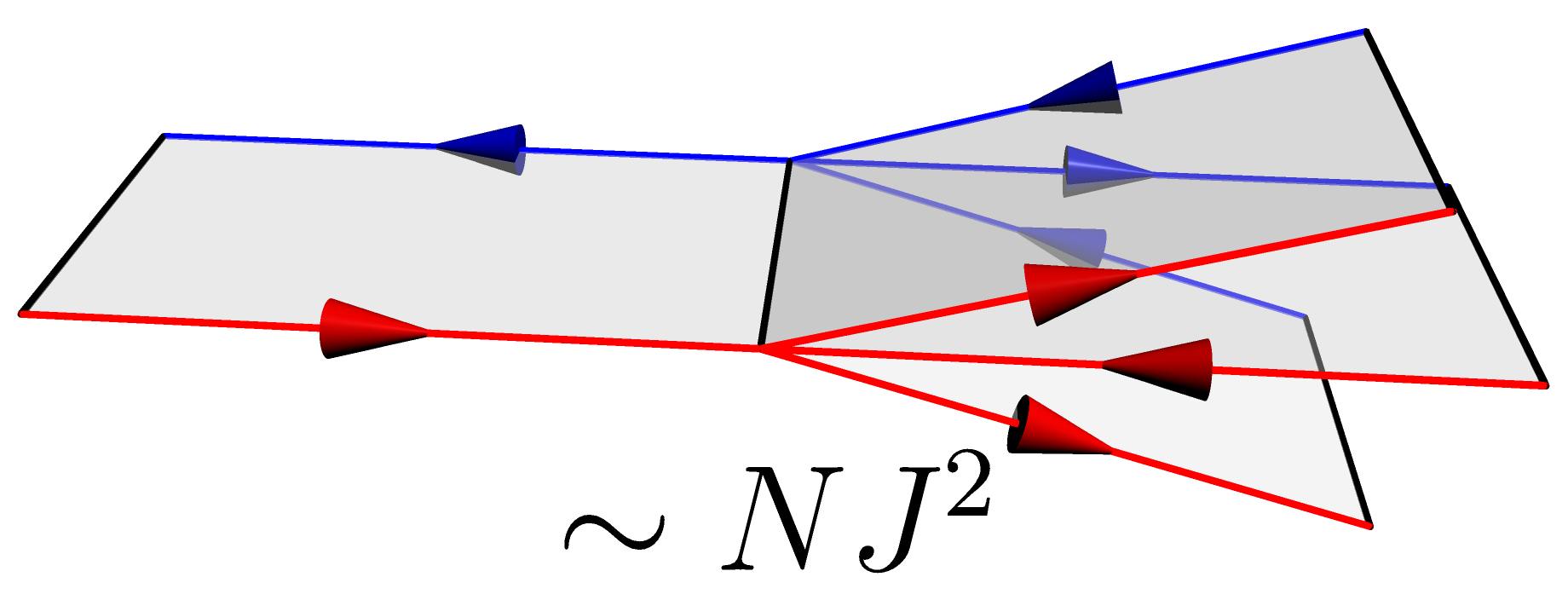}
  \caption{Interactions arise in an expansion around the $(G\,, \Sigma)$ saddle point from expanding the $\ln det (-\partial_\tau + \mu - \Sigma)$ term, which leads to arbitrary order sheets for which $\Sigma$ propagators can be attached to. Additionally, four $G$ fields can be attached at a ``seam.''}
  \label{fig:feynmanInts}
\end{figure}
The trace notation in the first term is shorthand for four time integrals, i.e. $\Tr [ G \Sigma] \equiv \int \dd{\tau_a} \dd{\tau_b} G(\tau_a, \tau_b) \Sigma(\tau_b, \tau_a)$. We can invert the quadratic action to obtain a propagator, which we can do separately for the replica diagonal and replica off-diagonal components. For the latter, we have
\begin{equation}
  \begin{aligned}
    -\frac{2 \delta S}{N} = \int \dd{\tau_{1, 2, 3, 4}}
  \begin{pmatrix} 
    \delta G^{ab}(\tau_1, \tau_2) & \delta \Sigma^{ab}(\tau_1, \tau_2) 
  \end{pmatrix} 
  \begin{pmatrix} 
    - t^2 \delta_{\tau_1, \tau_3} \delta_{\tau_2, \tau_4} & \delta_{\tau_1, \tau_3} \delta_{\tau_2, \tau_4}  \\
    \delta_{\tau_1, \tau_3} \delta_{\tau_2, \tau_4}  & -G(\tau_1 - \tau_3) G(\tau_2 - \tau_4)
  \end{pmatrix} 
    \begin{pmatrix} 
    \delta G^{ba}(\tau_4, \tau_3) \\ \delta \Sigma^{ba}(\tau_4, \tau_3) 
  \end{pmatrix} 
  \end{aligned}
\end{equation}
The matrix must be inverted, which can most easily be done in Matsubara frequency space. This leads to the result 
\begin{equation}
\begin{aligned}
    \langle \delta G^{a \neq b}(\tau_1, \tau_2) \delta G^{b \neq a}(\tau_4, \tau_3) \rangle = \frac{1}{N \beta^2}\sum_{i \omega_n, i \epsilon_n} e^{- i \omega_n (\tau_1 - \tau_3)- i \epsilon_n (\tau_4 - \tau_2)} \frac{G(i\omega_n) G(i \epsilon_n)}{1 - t^2 G(i \omega_n) G(i \epsilon_n)}\,,
    \\
    \langle \delta \Sigma^{a \neq b}(\tau_1, \tau_2) \delta G^{b \neq a}(\tau_4, \tau_3) \rangle = \frac{1}{N \beta^2}\sum_{i \omega_n, i \epsilon_n} e^{- i \omega_n (\tau_1 - \tau_3)- i \epsilon_n (\tau_4 - \tau_2)} \frac{1}{1 - t^2 G(i \omega_n) G(i \epsilon_n)}\,,
    \\
    \langle \delta \Sigma^{a \neq b}(\tau_1, \tau_2) \delta \Sigma^{b \neq a}(\tau_4, \tau_3) \rangle = \frac{1}{N \beta^2}\sum_{i \omega_n, i \epsilon_n} e^{- i \omega_n (\tau_1 - \tau_3)- i \epsilon_n (\tau_4 - \tau_2)} \frac{t^2}{1 - t^2 G(i \omega_n) G(i \epsilon_n)}\,.
  \end{aligned}
\end{equation}
This gives the expected result for $g_o$ in Eq.~\ref{eq:syk2ladders} of the main text once the trivial disconnected piece of $g_o$ is subtracted off. Note that for $t=0$, while $\langle \delta G^{ab} \delta G^{ba} \rangle$ is non-zero, its contribution to $g_o$ is subtracted off exactly by the disconnected piece. Hence, the leading order contribution to $g_o$ when $t=0$ is given by the first correction to the $G^{ab}$ propagator, illustrated in Fig.~\ref{fig:oDiagrams}. This corresponds to the fermionic Feynman diagram shown in the top of Fig.~\ref{fig:complexSYKConnected} in the main text.

\section{Statistics of ratio distributions}
\label{sec:ratioApp}
Here, we provide a summary of relevant results involving ratio distributions, which we utilize for calculating statistical properties of the thermopower.

We take $X_1$, $X_2$ to be two correlated Gaussian random variables, with means $\mu_{1, 2}$, variances $\sigma^2_{1, 2}$, and correlation coefficient $r$. Our quantity of interest is the random variable $Z \equiv X_1 / X_2$. The probability density function $f(z)$ of $Z$ can be obtained from the joint density $g(x_1, x_2)$ of $X_{1, 2}$,
\begin{equation}
  \begin{aligned}
    f(z) &= \int_{-\infty}^\infty \abs{y} g(z y, y) \dd{y} \,.
  \end{aligned}
\end{equation}
This function along with the cumulative distribution function $F(z) \equiv \int_{-\infty}^z f(x) \dd{x}$ are known~\cite{hinkley1969}. However, much like the Cauchy distribution - which is a limiting case of a ratio distribution when the numerator and denominator have zero mean - the integrals $\int_{-\infty}^\infty z^
a f(z) \dd{z}$, $\alpha \geq 1$ do not converge and the mean and variance are formally ill-defined.

One can make progress in the limit where $\abs{\sigma_2 / \mu_2} \rightarrow 0$; or in other words, when the probability of the denominator in $Z$ becoming negative is zero. This result can equivalently be derived from the assumption that $X_2 > 0$ which implies $F(z) \equiv P(x_1 / x_2 < z) = P(x_1 - z x_2 < 0)$. Since the sum of two correlated Gaussians is also a Gaussian, this gives the cumulative distribution function 
\begin{equation}
  \begin{aligned}
  F(z) = \Phi\left( \frac{\mu_2 z - \mu_1}{\sqrt{\sigma_1^2 - 2 z r \sigma_1 \sigma_2 + z^2 \sigma_2^2}} \right) 
  \end{aligned}
\end{equation}
where $\Phi(x)$ is the cumulative distribution function of a Gaussian random variable, $\Phi(x) \equiv \int_{-\infty}^x \phi(y) \dd{y}$, $\phi(x) \equiv \frac{1}{\sqrt{2\pi}} e^{-\frac{1}{2} x^2}$. 

For small fluctuations around the mean value, $\overline{z} = \mu_1 / \mu_2$, we have
\begin{equation}
  \begin{aligned}
  \Phi\left( \frac{\mu_2 z - \mu_1}{\sqrt{\sigma_1^2 - 2 z r \sigma_1 \sigma_2 + z^2 \sigma_2^2}} \right)  \approx \Phi \left( \frac{z - \overline{z}}{\overline{z} \sqrt{\frac{\sigma_1^2}{\mu_1^2} - \frac{2 r \sigma_1 \sigma_2}{\mu_1 \mu_2} + \frac{\sigma_2^2}{\mu_2^2}}} \right) 
  \end{aligned}
\end{equation}
which yields the approximation to normality, with variance
\begin{equation}
  \begin{aligned}
    \frac{\text{Var }z}{\overline{z}^2} \approx \frac{\sigma_1^2}{\mu_1^2}- \frac{2 r \sigma_1 \sigma_2}{\mu_1 \mu_2} + \frac{\sigma_2^2}{\mu_2^2}\,.
  \end{aligned}
\end{equation}

\begin{figure}[htpb]
  \centering
  \includegraphics[width=0.8\textwidth]{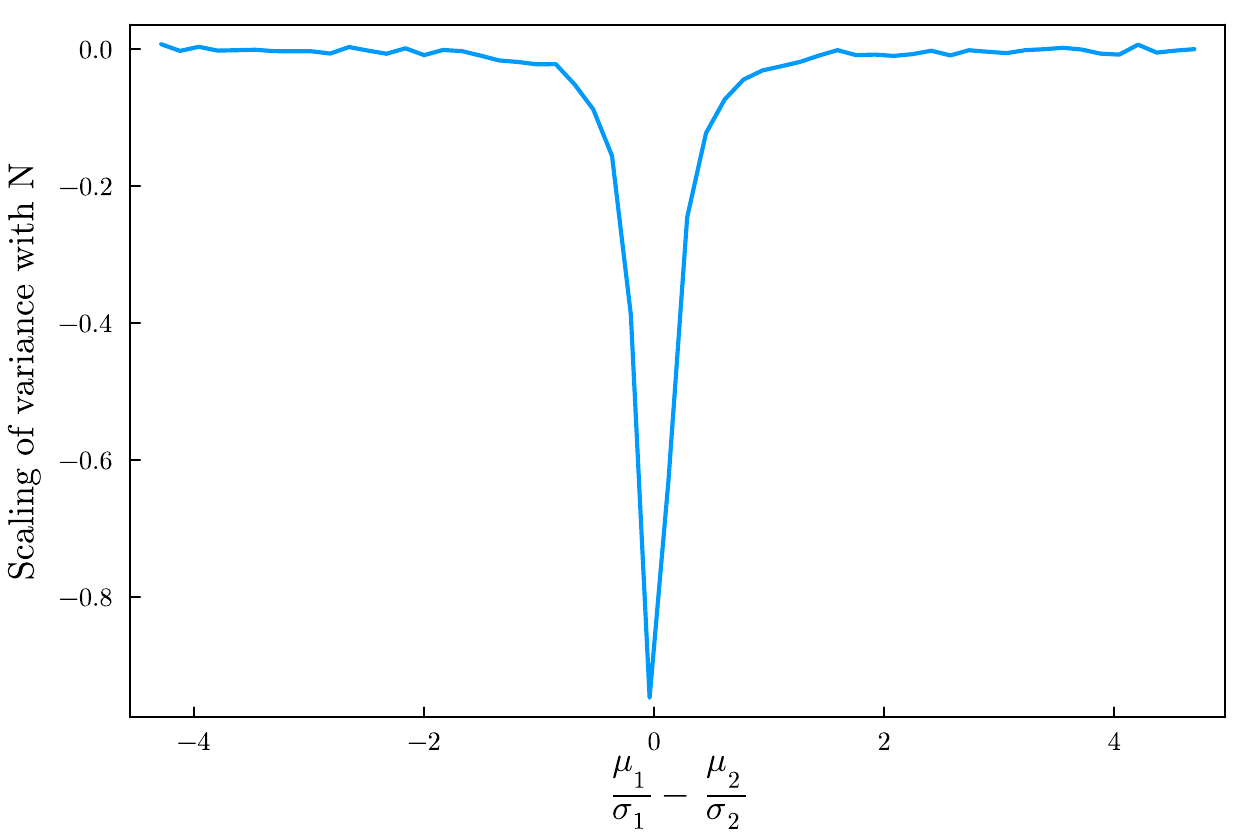}
  \caption{By drawing from a ratio distribution, where the correlation coefficient between the numerator and denominator is given by $1 - 1 / N$, we fit the variance to an $N^\alpha$ form and plot the exponent $\alpha$. When the probability distributions are tuned such that $\sigma_1 / \mu_1 = \sigma_2 / \mu_2$, we obtain a $N^{-1}$ suppression of the variance.}
  \label{fig:ratioScaling}
\end{figure}

In the main text, we find several situations where the numerator and denominator are highly correlated such that $r = 1 - \order{N^{-1}}$, where we use $N$ as a stand-in for a generic large dimensionless parameter, which depending on the context may refer to either the actual system size or $T / E_{\text{coh}}$. To leading order in $N^{-1}$, we therefore have \textit{perfect} correlation between the numerator and denominator, leading to 
\begin{equation}
  \begin{aligned}
    \frac{\text{Var }z}{\overline{z}^2} \approx \left(\frac{\sigma_1}{\mu_1}- \frac{\sigma_2}{\mu_2}\right)^2 + \order{N^{-1}}\,.
  \end{aligned}
\end{equation}
Working in the limit of perfect correlation means that we may think of $X_1$ and $X_2$ as arising from the same normal distribution $X$, i.e. $X_1 = \sigma_1  X + \mu_1$ and $X_2 = \sigma_2 X + \mu_2$. The ratio distribution is still non-trivial even if both variables arise from the same probability distribution. However, it does imply a special limit $\sigma_1 / \mu_1 = \sigma_2 / \mu_2$ where the distribution becomes trivial and the variance vanishes due to the numerator and denominator being directly proportional to each other. In this limit, the variance incurs an additional $N^{-1}$ suppression due to the necessity of expanding out $r$ to higher order. This prediction is confirmed by numerical simulation, see Fig.~\ref{fig:ratioScaling}. We take $10000$ samples of the ratio distribution $Z$ for parameters $\sigma_1^2 = 1.5$, $\sigma_2^2 = 1$, $\mu_y = 10$, $r = 1 - 1 / N$, and variable $\mu_x$. We fit the power law scaling of the variance as a function of $N$ for $500 < N < 10000$ and plot the exponent while varying $\mu_x$. As expected, an anomalous suppression of the variance appears at the critical value where $\sigma_1 / \mu_1 = \sigma_2 / \mu_2$.

\section{Conductance fluctuations for single-lead coupling}
\label{app:singleLead}
In the main text, we present results for conductance fluctuations for a model where we take our leads to be coupled to all sites with equal magnitude. To leading order in $N^{-1}$, fluctuations are controlled by the off-diagonal Green's function covariance $\rho_o$. If we instead choose to model our leads as only being coupled to a single site in the quantum dot, our results are modified as fluctuations are driven by the diagonal Green's function covariance $\rho_d$, which is generically suppressed relative to $\rho_o$ by an additional factor of $N^{-1}$. Note that this contribution is still present in our model in the main text, but is ignored in virtue of this $N^{-1}$ suppression. We present results for both $\rho_o$ and $\rho_d$ in the main text but focus on conductance fluctuations for fully-connected leads. Here, we present results for conductance fluctuations that arise from $\rho_d$, which are subleading in $N^{-1}$ for fully-connected leads but are the dominant contribution for leads coupled to a single site. We remind the reader that average conductance is insensitive to this choice and remains the same as in the main text.

For a free fermion model, we have the physical interpretation that $\rho_d$ gives the covariance of the single-particle eigenvalues, the form of which is universal and well-known from random matrix theory. In particular, the variance of linear statistics such as the conductance is given by the Dyson-Mehta formula~\cite{dyson1963, dyson1972}, which yields the conductance variance
\begin{equation}
  \begin{aligned}
    \text{Var } \sigma_{FF} = \left( \frac{e^2}{\hbar} \frac{\Gamma}{T N} \right)^2 \frac{3 \zeta(3)}{\pi^4}\,.
    \label{eq:freeFermionDM}
  \end{aligned}
\end{equation}
For a pure SYK model, our expression for $\rho_d$ given in Eq.~\ref{eq:sykrhod} yields
\begin{equation}
  \begin{aligned}
    \text{Var } \sigma_{SYK} = \left(\frac{e^2 \Gamma}{\hbar} \right) ^2  \frac{0.07}{N^4 J T}\,.
  \end{aligned}
\end{equation}
We now consider the case with both SYK interactions and random hopping terms. For the low temperature Fermi liquid phase, we predict a scaling similar to the free fermion result in Eq.~\ref{eq:freeFermionDM}, but with a renormalization which can be deduced on dimensional grounds to be
\begin{equation}
  \begin{aligned}
    \text{Var } \sigma_{tSYK} \propto \left( \frac{\Gamma e^2}{\hbar T N} \frac{t}{J}  \right)^2 \,, \quad T \ll E_{\text{coh}} \,.
  \end{aligned}
\end{equation}
For the SYK regime, $T \gg E_{\text{coh}}$, we find nearly identical to the case considered to the main text, due to the fact that in this regime, $\rho_d(\omega, \epsilon) = N^{-1} \rho_o(\omega, \epsilon)$ to leading order in $E_{\text{coh}} / T$. Hence,
\begin{equation}
  \begin{aligned}
    \text{Var } \sigma_{tSYK} = 2.02  \mathcal{E}^2 \left( \frac{\Gamma e^2}{\hbar N T } \frac{t}{J} \right)^2 \,, \quad T \gg E_{\text{coh}}\,.
  \end{aligned}
\end{equation}
Note that this is the same scaling as in the Fermi liquid regime, albeit with the crucial difference that the overall coefficient is proportional to the particle-hole asymmetry.
\bibliography{henryRefs.bib}
\end{document}